\documentclass[12pt]{article}
\usepackage{amssymb}

\def\hybrid{\topmargin 0pt      \oddsidemargin 0pt
        \headheight 0pt \headsep 0pt
        \voffset=-0.5cm
        \hoffset=-0.25in
        \textwidth 6.75in
        \textheight 9.5in       
        \marginparwidth 0.0in
        \parskip 5pt plus 1pt   \jot = 1.5ex}
\catcode`\@=11
\def\marginnote#1{}

\newcount\hour
\newcount\minute
\newtoks\amorpm
\hour=\time\divide\hour by60 \minute=\time{\multiply\hour by60
\global\advance\minute by-\hour}
\edef\standardtime{{\ifnum\hour<12 \global\amorpm={am}%
        \else\global\amorpm={pm}\advance\hour by-12 \fi
        \ifnum\hour=0 \hour=12 \fi
        \number\hour:\ifnum\minute<10 0\fi\number\minute\the\amorpm}}
\edef\militarytime{\number\hour:\ifnum\minute<10 0\fi\number\minute}

\def\draftlabel#1{{\@bsphack\if@filesw {\let\thepage\relax
   \xdef\@gtempa{\write\@auxout{\string
      \newlabel{#1}{{\@currentlabel}{\thepage}}}}}\@gtempa
   \if@nobreak \ifvmode\nobreak\fi\fi\fi\@esphack}
        \gdef\@eqnlabel{#1}}
\def\@eqnlabel{}
\def\@vacuum{}
\def\draftmarginnote#1{\marginpar{\raggedright\scriptsize\tt#1}}
\def\draftlabel#1{{\@bsphack\if@filesw {\let\thepage\relax
   \xdef\@gtempa{\write\@auxout{\string
      \newlabel{#1}{{\@currentlabel}{\thepage}}}}}\@gtempa
   \if@nobreak \ifvmode\nobreak\fi\fi\fi\@esphack}
        \gdef\@eqnlabel{#1}}
\def\@eqnlabel{}
\def\@vacuum{}
\def\draftmarginnote#1{\marginpar{\raggedright\scriptsize\tt#1}}

\def\draft{\oddsidemargin -.5truein
        \def\@oddfoot{\sl preliminary draft \hfil
        \rm\thepage\hfil\sl\today\quad\militarytime}
        \let\@evenfoot\@oddfoot \overfullrule 3pt
        \let\label=\draftlabel
        \let\marginnote=\draftmarginnote
   \def\@eqnnum{(\theequation)\rlap{\kern\marginparsep\tt\@eqnlabel}%
\global\let\@eqnlabel\@vacuum}  }

\def\numberbysection{\@addtoreset{equation}{section}
        \def\theequation{\thesection.\arabic{equation}}}

\def\underline#1{\relax\ifmmode\@@underline#1\else
        $\@@underline{\hbox{#1}}$\relax\fi}

\def\titlepage{\@restonecolfalse\if@twocolumn\@restonecoltrue\onecolumn
     \else \newpage \fi \thispagestyle{empty}\c@page\z@
        \def\thefootnote{\fnsymbol{footnote}} }

\def\endtitlepage{\if@restonecol\twocolumn \else  \fi
        \def\thefootnote{\arabic{footnote}}
        \setcounter{footnote}{0}}  

\numberbysection \hybrid

\newcounter{mo}

\newcommand{\tr}{{\rm tr}}

\newcommand{\vf}{\varphi}
\newcommand{\al}{\alpha}
\newcommand{\be}{\beta}
\newcommand{\ga}{\gamma}
\newcommand{\om}{\omega}
\newcommand{\vth}{\vartheta}

\newcommand{\Mat}{ {\rm Mat}(N,\mathbb C) }

\newcommand{\mC}{\mathbb C}
\newcommand{\mZ}{\mathbb Z}

\newtheorem{predl}{Proposition}[section]
\newtheorem{example}{Example}[section]

\def\beq{\begin{equation}}
\def\eq{\end{equation}}
\def\p{\partial}

\newtheorem{theor}{Theorem}

\newcommand{\mat}[4]{\left(\begin{array}{cc}{#1}&{#2}\\ \ \\{#3}&{#4}
\end{array}\right)}

\def\res{\mathop{\hbox{Res}}\limits}

\begin{document}

\setcounter{page}{1}

\date{}
\date{}


\

\begin{center}
%
 {\LARGE{On factorized Lax pairs for classical many-body}}
 \\ \ \\
  {\LARGE{integrable systems}}

\vspace{15mm}

{\large  {M. Vasilyev}\,\footnote{Steklov Mathematical Institute of
Russian Academy of Sciences, 8 Gubkina St., Moscow 119991, Russia;
e-mail: mikhail.vasilyev@phystech.edu}
 \quad\quad\quad\quad\quad
 {A. Zotov}\,\footnote{Steklov Mathematical Institute of Russian Academy of Sciences, 8 Gubkina St., Moscow 119991, Russia;
  e-mail:
zotov@mi.ras.ru}
 }
\end{center}

\vspace{5mm}


 \begin{abstract}
In this paper we study factorization formulae for the Lax matrices
of the classical Ruijsenaars-Schneider and Calogero-Moser models. We
review the already known results and discuss their possible origins.
The first origin comes from the IRF-Vertex relations and the
properties of the intertwining matrices. The second origin is based
on the Schlesinger transformations generated by modifications of
underlying vector bundles. We show that both approaches provide
explicit formulae for $M$-matrices of the integrable systems in
terms of the intertwining matrices (and/or modification matrices).
In the end we discuss the Calogero-Moser models related to classical
root systems. The factorization formulae are proposed for a number
of special cases.
 \end{abstract}

\vspace{2mm}

 \footnotesize \tableofcontents  \normalsize




\section{Introduction}
\setcounter{equation}{0}
In this paper we deal with the Lax pairs of the Calogero-Moser
\cite{Ca1,Ca2,Ca3,Ca4,Ca5}, \cite{Krich1} and Ruijsenaars-Schneider
\cite{Ruijs11,Ruijs12} models. More precisely, we study the
factorization formulae for the Lax matrices of these models. For the
elliptic ${\rm gl}_N$ Ruijsenaars-Schneider model it is of the
form\footnote{The form (\ref{q001}) is defined up to multiplication
of $L(z)$ by a scalar non-dynamical function. In what follow we will
fix this freedom as given in (\ref{q51}) to match the custom form
(\ref{q022}).}:
  \beq\label{q001}
  \begin{array}{l}
  \displaystyle{
 L^{\rm RS}(z)=g^{-1}(z)g(z+\hbar')e^{P/c}\in\Mat\,,
 }
 \end{array}
 \eq
where $\hbar'$ and $c$ are constants, $z$ is the spectral parameter,
and
  \beq\label{q002}
  \begin{array}{l}
  \displaystyle{
  P={\rm diag}(p_1,...,p_N)\in\Mat\,,\quad\quad
  g(z)=g(z,q_1,...,q_N)\in\Mat\,,
 }
 \end{array}
 \eq
 where $g(z,q)$ is given by (\ref{q52}).
The positions of particles $q_i$ and momenta $p_i$ are canonically
conjugated $\{p_i,q_j\}=\delta_{ij}$. The form (\ref{q001}) was
observed in \cite{Hasegawa1,Hasegawa2,Hasegawa3} at quantum level.
It was used for the proof that the quantum version of the gauge
transformed Lax matrix
  \beq\label{q003}
  \begin{array}{l}
  \displaystyle{
 g(z)L^{\rm RS}(z)g^{-1}(z)=g(z+\hbar')e^{P/c}g^{-1}(z)
 }
 \end{array}
 \eq
satisfies the quantum exchange (or $RLL$) relations with the
non-dynamical Baxter-Belavin $R$-matrix
\cite{Baxter,Belavin}\footnote{The expression (\ref{q003}) itself
satisfies the classical quadratic exchange relations (\ref{q320})
with the classical (non-dynamical) $r$-matrix of the
Belavin-Drinfeld \cite{BD} type (in the elliptic case).}. In $N=2$
case this result reproduces the representation of the quantum
Sklyanin algebra \cite{Skl1} through the difference operators
\cite{Skl2}, and for generic $N$ it provides similar representation
for the ${\rm GL}_N$ analogue of the Sklyanin algebra
\cite{Cherednik}.  The application to exchange relations establish a
link between (\ref{q001}) and the IRF-Vertex correspondence
\cite{Baxter2,Jimbo1,Jimbo2,Jimbo3}, which maps dynamical and
non-dynamical $R$-matrices into each other. Up to some additional
diagonal gauge the matrix $g(z)$ entering (\ref{q001}) is the matrix
of the intertwining vectors introduced (for the elliptic case) in
\cite{Baxter2,Jimbo1,Jimbo2,Jimbo3}. It is used for construction of
the elliptic analogue of the Drinfeld twist \cite{Dr1,Dr11}. We will
review the above mentioned relations in the next Section. The
classical analogue of the IRF-Vertex relations based on (\ref{q001})
and the corresponding parameterization  of the classical Sklyanin
algebra in terms of the Ruijsenaars-Schneider variables (the
classical bosonization formulae or the classical representation
formulae) are directly obtained from the results of
\cite{Hasegawa1,Hasegawa2,Hasegawa3}. See \cite{Chen11,Chen12} for
the quasi-classical limit. A general form for such parameterization
follows from (\ref{q003}) by taking residue at $z=0$. Namely, the
components of the matrix
  \beq\label{q004}
  \begin{array}{l}
  \displaystyle{
 S=S(p,q,\hbar',c)=\res\limits_{z=0}\left(g(z)L^{\rm RS}(z)g^{-1}(z)\right)=
 g(\hbar')\,e^{P/c}\,{\breve g}(0)\,,\quad\quad {\breve
 g}(0)=\res\limits_{z=0}\,g^{-1}(z)
 }
 \end{array}
 \eq
are the generators of the classical Sklyanin algebra. In
(\ref{q004}) we also used the property of $g(z)$ that near $z=0$
 \beq\label{q005}
 \begin{array}{c}
  \displaystyle{
 g^{-1}(z)=\frac1z\,{\breve g}(0)+A+O(z)\,,
 }
 \end{array}
 \eq
 i.e. $g(z)$ is degenerated at $z=0$, and $\det g(z)$ has the first
 order zero at $z=0$. Let us also mention that the first example of
 the classical IRF-Vertex like relation was observed in \cite{ZaT}
 between the nonlinear Schr\"odinger equation and the classical
 Heisenberg magnet.

While in the elliptic case we deal with the Lax representation with
spectral parameter, for the trigonometric and rational cases there
are Lax representations without spectral parameter. The
factorization formulae exist for each of the cases. From the
IRF-Vertex relations viewpoint the trigonometric case without
spectral parameter is related to $R$-matrix structure of the chiral
Potts model \cite{Bazhanov1,Bazhanov2} based on
\cite{Chered,Kulish}, while the trigonometric case with spectral
parameter is described by the intertwining matrix of the
"non-standard" trigonometric $R$-matrix \cite{AHZ} generalizing the
7-vertex $R$-matrix \cite{Chered}. Similarly, the rational case
without spectral parameter is related to the $R$-matrices of the
Cremmer-Gervais type \cite{CG1,CG2,BFeher1,BFeher2,BFeher3}, while
the rational case with spectral parameter comes from ${\rm GL}_N$
generalization \cite{AASZ,LOZ8} of the rational 11-vertex $R$-matrix
\cite{Chered}. Factorization formulae for all the cases will be also
reviewed in the next Section.

In the non-relativistic limit $\hbar'=\nu'/c$, $c\rightarrow\infty$
(\ref{q001}) provides the Lax matrix of the Calogero-Moser model
written in the following form:
  \beq\label{q006}
  \begin{array}{l}
  \displaystyle{
 L^{\rm CM}(z)=P+\nu'g^{-1}g'\,.
 }
 \end{array}
 \eq
where $g'=\p_zg(z)$ and $\nu'$ is the coupling constant. The custom
form of the elliptic model is achieved by setting $\nu'=N\nu$, see
(\ref{q0297}). Similarly to (\ref{q003}) the gauge transformed Lax
matrix
  \beq\label{q007}
  \begin{array}{l}
  \displaystyle{
 g(z)L^{\rm CM}(z)g^{-1}(z)=g(z)Pg^{-1}(z)+\nu'g'(z)g^{-1}(z)\,.
 }
 \end{array}
 \eq
satisfies the classical linear exchange relations (\ref{q315}) with
the classical (non-dynamical) $r$-matrix of the Belavin-Drinfeld
type (in the elliptic case). While the residue of (\ref{q003}) is
the classical representation of the Sklyanin algebra, the residue of
(\ref{q007})
  \beq\label{q008}
  \begin{array}{l}
  \displaystyle{
 S=S(p,q,\nu')=g(0)P{\breve g}(0)+\nu'g'(0){\breve g}(0)
 }
 \end{array}
 \eq
is the classical representation of the ${\rm gl}_N$ Lie algebra. The
Poisson brackets between the matrix elements of $S$ are the
Poisson-Lie brackets on the Lie coalgebra ${\rm gl}_N^*$. Moreover,
the matrix ${\breve g}(0)$ is of rank one (see (\ref{q362})), and
therefore, the $S$ matrices (\ref{q008}) and (\ref{q004}) are of
rank one too: $S=\xi(p,q)\otimes \psi(q)$. In the rational case  the
components of $\psi$ vector are elementary symmetric functions of
the coordinates $q_i$, while the components of the $\xi$ vector  are
canonically conjugated to those of $\psi$:
$\{\xi_i,\psi_j\}=\delta_{ij}$ (for the non-relativistic case
(\ref{q008})). These type variables were used for reformulation of
the quantum Calogero-Moser model in terms of the Lie algebra data in
\cite{Turbiner} and \cite{Perelomov1,Perelomov2}.



A general scheme for the classical IRF-Vertex relations was
suggested in \cite{LOZ} and is known as the symplectic Hecke
correspondence. It unifies  a set of integrable models related by
gauge transformations of $g(z,q)$ type. The Lax matrices under
consideration are known \cite{Krich1} to be sections of bundles over
the base spectral curve $\Sigma$ with a local coordinate $z$:
$L(z)\in\Gamma({\rm End}V,\Sigma)$. The underlying vector bundles
$V$ are also related by the action of the gauge transformations,
which change the degrees of the bundles by one. It happens due to
the special local structure (\ref{q005}) of $g(z,q)$. Its action
adds a zero (or a pole) towards a certain direction. Such gauge
transformations are called modifications of bundles
\cite{Dr2,Arinkin1,Arinkin2}. In this respect (\ref{q001}) is a
combination of two modifications \cite{Vakul}. The set of models
unified by the symplectic Hecke correspondence consists of the
Calogero-Moser model (including its spin generalizations), elliptic
integrable tops and intermediate models, which are described by
partially dynamical $R$-matrices
 \cite{LOSZ4,SZ}. The gauge transformation relating (\ref{q006}) and (\ref{q007}) is then treated
 as transition from the Calogero-Moser model (with variables $p_i,q_j$) to the special elliptic
 top, where the matrix of dynamical variables $S$ (\ref{q008})
 belong to the coadjoint orbit (of ${\rm GL}_N$ Lie group) of
the minimal dimension, i.e. when $S$ is of rank one. The relation
(\ref{q008}) provides explicit change of variables between the
systems in this case.

{\bf The purpose of the paper} is two-fold. The first one is to
clarify possible origins of the factorization formulae (\ref{q001})
and (\ref{q006}). In fact, the factorization is neither necessary
nor sufficient for integrability. A natural set up of the problem is
as follows.
Which
$g(z,q)$ provide the Lax matrices for integrable models? Put it
differently, for which $g(z,q)$ there exist $M$-matrix such that the
Lax equations
  \beq\label{q009}
  \begin{array}{l}
  \displaystyle{
  {\dot L}(z)=[L(z),M(z)]
 }
 \end{array}
 \eq
hold true identically in $z$ and are equivalent to equations of
motion of an integrable system defined by the Lax matrix
(\ref{q001}) or (\ref{q006})? It is easy to verify that a generic
matrix $g(z,q)$ does not provide Lax matrix.
Only very special $g(z,q)$ lead to an integrable system, and the
information about integrability of (\ref{q001}) or (\ref{q006}) is
encoded in the form of the matrix $g(z,q)$.
Therefore, it is reasonable to expect that the rest of the data (not
only the Lax matrix) is formulated through $g(z,q)$. We focus on
derivation of the $M$-matrices\footnote{$M$-matrices for all flows
can be found in terms of the classical $r$-matrix structures
\cite{Skl1,STS,ABT1,ABT2}. Here we deal with the $M$-matrices
related to the first non-trivial flows. On one hand, it is enough
for integrability. On the other hand, we are going to use the
monodromy preserving equations which correspond to these concrete
flows.} for the Calogero-Moser and Ruijsenaars-Schneider models in
terms of $g(z,q)$.

From the above we see that there are two natural possible origins
for $g(z,q)$ with the property that it provides Lax matrix of an
integrable model. They come from the algebraic and geometric
viewpoints. The algebraic origin is the IRF-Vertex correspondence,
i.e. the treatment of the matrix $g(z,q)$ as an intertwining matrix
(in the fundamental representation) entering the Drinfeld twist. The
geometric origin is interpretation of $g(z,q)$ matrix as
modification of bundle on the base spectral curve related to the Lax
matrix (\ref{q001}) or (\ref{q006}). Using these two treatments of
$g(z,q)$ we obtain expressions for the $M$-matrices of the
Ruijsenaars-Schneider (\ref{q001}) and Calogero-Moser (\ref{q006})
models. Namely, we prove the following
 \begin{theor}\label{theor}
 The M-matrix of the Ruijsenaars-Schneider model defined by the Lax matrix
 (\ref{q001}) can be written in terms of the $g(z,q)$ matrix
 (\ref{q52}) as follows:
   \beq\label{q010}
  \begin{array}{c}
  \displaystyle{
 {M^{\rm RS}}(z)=-g^{-1}(z)g'(z)G-F+g^{-1}(z)\frac{d}{dt}\,{ g}(z)\,,
 }
 \end{array}
 \eq
 with
   \beq\label{q011}
  \begin{array}{c}
  \displaystyle{
 G=\tr_2\left({\mathcal O}_{12}\frac{\vth'(0)}{\vth(\hbar)}\,{\breve
 g}_2(0)g_2(N\hbar)\,
 e^{P_2/c}\right)\,,\quad
 F=\tr_2\left({\mathcal O}_{12}\frac{\vth'(0)}{\vth(\hbar)}\,A_2\,g_2(N\hbar)\,
 e^{P_2/c}\right).
 }
 \end{array}
 \eq
 where we assume that (in the elliptic case) $\hbar'=N\hbar$, the
 matrix $A$ is the one from the expansion (\ref{q005}), and
    \beq\label{q012}
  \begin{array}{c}
  \displaystyle{
 \mathcal O_{12}=\sum\limits_{i,j}E_{ii}\otimes
E_{ji}\,.
 }
 \end{array}
 \eq
 \end{theor}
and
 \begin{theor}
 The M-matrix of the Calogero-Moser model defined by the Lax matrix
 (\ref{q006}) can be written in terms of the $g(z,q)$ matrix
 (\ref{q52}) as follows:
  \beq\label{q013}
  \begin{array}{c}
  \displaystyle{
 M=g^{-1}(z)\frac{d}{d\tau}\,g(z)-g^{-1}(z)\frac{d}{dt}\,g(z)
 }
 \end{array}
 \eq
where
  \beq\label{q014}
  \begin{array}{c}
  \displaystyle{
 {\rm diag}(q)_\tau-{\rm diag}(q)_t=-\frac1N\,d\,,\quad d_i=\sum\limits_{k\neq i}E_1(q_{ik})\,.
 }
 \end{array}
 \eq
 \end{theor}
The statements of both theorems hold true for trigonometric and
rational cases as well. The partial derivative with respect to the
moduli $\tau$ should be transformed into the second derivative with
respect to the argument (through the heat equation) in these cases.
See Section \ref{Sect35}.

The proof of the first statement (\ref{q010}) is based on the
algebraic treatment of $g(z,q)$. Following \cite{SeZ} we mention
that the IRF-Vertex correspondence provides the following relation
between quantum non-dynamical $R$-matrix and the intertwining matrix
$g(z,q)$:
  \beq\label{q015}
  \begin{array}{c}
  \displaystyle{
 \frac1N\,{\breve g}_2(0,q)\,R^\hbar_{12}(z)=g_1(z+N\hbar,q)\,
 \mathcal O_{12}\, g_2^{-1}(N\hbar,q)\,
 g_1^{-1}(z,q)\,,
 }
 \end{array}
 \eq
where $\mathcal O_{12}$ is (\ref{q012}). Next, we use the $R$-matrix
formulation for integrable tops based on the quasiclassical limit of
1-site chain \cite{Skl1}. It was shown in \cite{LOZ16} that the Lax
equations (\ref{q009}) with
  \beq\label{q016}
  \begin{array}{c}
  \displaystyle{
L^\hbar(S,z)=\frac1N\,\tr_2\left(R_{12}^\hbar(z)S_2\right)\,,\quad\quad
M^\hbar(S,z)=-\frac1N\,\tr_2\left(r_{12}(z)S_2\right),
 }
 \end{array}
 \eq
 where $r_{12}(z)$ is the classical $r$-matrix ($R_{12}^\hbar(z)=1\otimes 1\hbar^{-1}+r_{12}(z)+O(\hbar)$),
provide equations of motion for the (relativistic) top model if the
quantum unitary $R$-matrix satisfies the associative Yang-Baxter
equation.
It is verified explicitly using (\ref{q015}) that under substitution
$S=S(p,q,\hbar,c)$ (\ref{q004}) the Lax matrix $L^\hbar(S,z)$ turns
into the gauged transformed Ruijsenaars-Schneider one (\ref{q003}).
Therefore, the $M$-matrix of the Ruijsenaars-Schneider model can be
evaluated by the inverse gauge transformation of the
$M^\hbar(S(p,q,\hbar,c),z)$. In this way we come to the expression
(\ref{q010}), which is then verified by direct calculation.

The proof of the second Theorem (\ref{q013}) uses the geometric
treatment of $g(z,q)$. The non-trivial part of the Lax matrix
(\ref{q006}) is a $z$-component of the pure gauge connection. To
obtain it we need to allow transition from the Lax matrix to the
connection along the $z$ coordinate on the base spectral curve. It
is exactly the statement of the Painlev\'e-Calogero correspondence
\cite{LO}: the Lax pair of the elliptic Calogero-Moser model
satisfies not only the Lax equation (\ref{q009}) but also the
monodromy preserving equations (zero-curvature condition)
  \beq\label{q017}
  \begin{array}{c}
  \displaystyle{
 2\pi\imath\frac{d}{d\tau}L-\frac{d}{dz}M=[L,M]\,,
 }
 \end{array}
 \eq
which lead to the higher Painlev\'e equations (\ref{q703}) with the
time variable being the moduli of the elliptic curve $\tau$. Then
the Lax matrix (\ref{q006}) can be obtained by combining the
Schlesinger transformation (the action of the modification of bundle
on the connection) and the Painlev\'e-Calogero correspondence, see
(\ref{q705}). Applying the same procedure to the $M$-matrix we come
to the from (\ref{q013}).

Another purpose of the paper is to study possible extension of the
factorization formulae to the models associated with the root
systems of the classical Lie algebras
\cite{OP1,OP2,ABT1,DP,BCS1,BCS2,Feher11,Feher12,Feher13,Feher14,Chen21,Chen22,Chen23}.
Some of the constructions discussed above are naturally extended to
these cases. For instance, the symplectic Hecke correspondence and
underlying modifications of bundles can be defined for $G$-bundles
with $G$ being a simple complex Lie group \cite{LOSZ1,LOSZ2}. At the
same time the intertwining matrix in the elliptic case is known to
exist for $A_N$ root system only \cite{BD}. The question which
intertwining vectors generate the factorized Lax pairs deserves
further elucidations.

Instead of using (\ref{q001}) and/or (\ref{q006}) in the rational
(and trigonometric) cases without spectral parameter we can rewrite
them in a slightly different way using that $g'=C_0g$ in these
cases, where $C_0$ is some constant matrix. This is due to
$g$-matrix for the latter cases is of Vandremonde type. Then
(\ref{q006}) turns into
  \beq\label{q018}
  \begin{array}{l}
  \displaystyle{
 L^{\rm CM}(z)=P+\nu'g^{-1}C_0g\,.
 }
 \end{array}
 \eq
 In the last Section we propose factorization formulae of type (\ref{q018}) for
 the rational Calogero models related to root systems $B,C,D$. This
 study is inspired by possible application to quantum-classical
 duality \cite{GZZ1,GZZ2,GZZ3}.


\section{Brief review}\label{sect2}
\setcounter{equation}{0}

\subsection{Ruijsenaars-Schneider and Calogero-Moser models}
The elliptic ${\rm gl}_N$ Ruijsenaars-Schneider model
\cite{Ruijs11,Ruijs12} describes $N$ interacting particles on the
complex plane with positions $q_k$ and equations of motion
  \beq\label{q021}
  \begin{array}{c}
  \displaystyle{
 {\ddot q}_i=\sum\limits_{k\neq i}^N{\dot q}_i{\dot q}_k
 (2E_1(q_{ik})-E_1(q_{ik}+\hbar)-E_1(q_{ik}-\hbar))\,,\quad i=1\dots
 N\,,
 }
 \end{array}
 \eq
 where $q_{ij}=q_i-q_j$, $E_1(x)$ is the function (\ref{q912}) and
 $\hbar$ is the coupling constant. The model is described
 by $\Mat$-valued Lax matrix with the spectral parameter $z$:
  \beq\label{q022}
  \begin{array}{l}
  \displaystyle{
 L^{\rm RS}_{ij}=\phi(z,q_{ij}+\hbar)\,\prod_{k\neq j}\frac{\vth(q_{jk}-\hbar)}{\vth(q_{jk})}\,e^{p_j/c}=
 \phi(z,q_{ij}+\hbar)\,\frac{D_j^{-\hbar}}{D_j^0}\,e^{p_j/c}\,,\ \
 D_j^\eta=\prod_{k\neq j}{\vth(q_{jk}+\eta)}\,,
 }
 \end{array}
 \eq
where $c$ is the light speed and $\phi(x,y)$ is the Kronecker
function (\ref{q911}). The Hamiltonian arises as the trace of
(\ref{q022}). More precisely,
  \beq\label{q023}
  \begin{array}{l}
  \displaystyle{
 H^{\rm RS}=c\frac{\tr L^{\rm RS}}{\phi(z,\hbar)}=c\sum\limits_{j=1}^N \frac{D_j^{-\hbar}}{D_j^0}\,e^{p_j/c}\,.
 }
 \end{array}
 \eq
 Then\footnote{The canonical Poisson brackets are assumed: $\{p_i,q_j\}=\delta_{ij}$.}
  \beq\label{q024}
  \begin{array}{l}
  \displaystyle{
 {\dot q}_j=\frac{D_j^{-\hbar}}{D_j^0}\,e^{p_j/c}\,.
 }
 \end{array}
 \eq
and the Lax matrix (\ref{q022}) acquires the form:
  \beq\label{q025}
  \begin{array}{l}
  \displaystyle{
 L^{\rm RS}_{ij}=\phi(z,q_{ij}+\hbar)\,{\dot q}_j\,.
 }
 \end{array}
 \eq
The definition of the velocities (\ref{q024}) is not unique.  A
family of canonical maps
  \beq\label{q026}
  \begin{array}{c}
  \displaystyle{
p_j\ \rightarrow\ p_j+c_1\log \prod\limits_{k\neq
j}^N\frac{\vth(q_{j}-q_k+c_2)}{\vth(q_{j}-q_k-c_2)}\,,
 }
 \end{array}
 \eq
with arbitrary constants $c_{1,2}$ can be used as well. Equations of
motion (\ref{q021}) (they are independent of (\ref{q026})) can be
written in the Lax form
  \beq\label{q027}
  \begin{array}{c}
  \displaystyle{
 {\dot L}^{\rm RS}\equiv\{H^{\rm RS},L^{\rm RS}\}=[L^{\rm RS},M^{\rm
 RS}]\,,
 }
 \end{array}
 \eq
 where the $M$-matrix is as follows:
  \beq\label{q028}
  \begin{array}{c}
  \displaystyle{
 M^{\rm RS}_{ij}=-(1-\delta_{ij})\phi(z,q_i-q_j)\,{\dot q}_j
 }
 \\ \ \\
  \displaystyle{
 -\delta_{ij}\Big({\dot q}_i\,(E_1(z)+E_1(\hbar)) +\sum\limits_{k\neq i} {\dot q}_k\,(E_1(q_{ik}+\hbar)-E_1(q_{ik})) \Big)
 \,.
 }
 \end{array}
 \eq
In the non-relativistic limit $\hbar=\nu/c$, $c\rightarrow\infty$
the Lax pair of the Calogero-Moser model \cite{Ca1,Ca2,Ca3,Ca4,Ca5}
is reproduced \cite{Krich1}:
  \beq\label{q029}
  \begin{array}{c}
  \displaystyle{
 L^{\rm CM}_{ij}=({\dot q}_i+\nu E_1(z))\,\delta_{ij}+\nu(1-\delta_{ij})\phi(z,q_{ij})\,,\quad\quad
 {\dot q}_i=p_i-\nu\sum\limits_{k\neq i}E_1(q_{ik})\,,
 }
 \end{array}
 \eq
  \beq\label{q030}
  \begin{array}{c}
  \displaystyle{
M^{\rm CM}_{ij}=\nu d_i\,\delta_{ij}
+\nu(1-\delta_{ij})f(z,q_{ij})\,,\quad d_i=\sum\limits_{k\neq i}
E_2(q_{ik})\,,
 }
 \end{array}
 \eq
See the definitions of $E_2(x)$ and $f(x,y)$ in (\ref{q912}),
(\ref{q917}). The Hamiltonian
  \beq\label{q031}
  \begin{array}{c}
  \displaystyle{
 H^{\hbox{\tiny{CM}}}=\sum\limits_{i=1}^N\frac{{\dot q}_i^2}{2}-\nu^2\sum\limits_{i>j}^N\wp(q_i-q_j)\,,
 }
 \end{array}
 \eq
where ${\dot q}_i={\dot q}_i(p,q)$ (\ref{q029}) provides equations
of motion
  \beq\label{q032}
  \begin{array}{c}
  \displaystyle{
 {\ddot q}_i=\nu^2\sum\limits_{k\neq i}\wp'(q_{ik})\,.
 }
 \end{array}
 \eq
In trigonometric and rational cases the functions used above are as
follows. In the trigonometric limit
  \beq\label{q033}
  \begin{array}{c}
  \displaystyle{
 \phi(z,q)\rightarrow \coth(z)+\coth(q)\,,\quad\quad
 E_1(z)\rightarrow \coth(z)\,,\quad D_j^\eta\rightarrow\prod_{k\neq j}{\sinh(q_{jk}+\eta)}\,,
 }
\\
  \displaystyle{
 f(z,q)\rightarrow -\frac{1}{\sinh^2(q)}\,,\quad\quad
 E_2(z)\,, \wp(z)\rightarrow \frac{1}{\sinh^2(z)}\,,
 }
 \end{array}
 \eq
and in the rational limit
  \beq\label{q034}
  \begin{array}{c}
  \displaystyle{
 \phi(z,q)\rightarrow \frac1z+\frac1q\,,\quad\quad
 E_1(z)\rightarrow \frac1z\,,\quad\quad D_j^\eta\rightarrow\prod_{k\neq
 j}{(q_{jk}+\eta)}\,,
 }
 \\
  \displaystyle{
 f(z,q)\rightarrow-\frac{1}{q^2}\,,\quad\quad
  E_2(z)\,, \wp(z)\rightarrow\frac{1}{z^2}\,.
 }
 \end{array}
 \eq

\subsection{Elliptic integrable tops}
The elliptic top \cite{LOZ} is the model of the Euler-Arnold type.
Dynamical variables are arranged into matrix $S\in\Mat$, and the
equations of motion are
 \beq\label{q310}
 \begin{array}{c}
  \displaystyle{
\dot S=[S,J(S)]\,,\quad S=\sum\limits_{i,j=1}^N
E_{ij}S_{ij}=\sum\limits_{\al\in\,\mZ_N\times\mZ_N;\, \al\neq
0}T_\al S_\al\,,
 }
 \end{array}
 \eq
 \beq\label{q311}
 \begin{array}{c}
  \displaystyle{
  J(S)=\sum\limits_{\al\neq 0}T_\al S_\al J_\al\,,\quad J_\al=-E_2(\om_\al)\,,\quad
 \om_\al=\frac{\al_1+\al_2\tau}{N}\,,
 }
 \end{array}
 \eq
where $\{E_{ij}\}$ is the standard matrix basis and $\{T_\al\}$ is
the one (\ref{q901}). The Lax pair is of the form:
 \beq\label{q312}
 \begin{array}{c}
  \displaystyle{
L^{\rm top}(z)=\sum\limits_{\al\neq 0}T_\al
S_\al\vf_\al(z,\om_\al)\,,\quad M^{\rm top}(z)=\sum\limits_{\al\neq
0}T_\al S_\al f_\al(z,\om_\al)\,.
 }
 \end{array}
 \eq
The Hamiltonian
 \beq\label{q313}
 \begin{array}{c}
  \displaystyle{
H^{\rm top}=\sum\limits_{\al\neq 0}S_\al S_{-\al}E_2(\om_\al)
 }
 \end{array}
 \eq
 is evaluated from $\tr(L^{\rm top}(z))^2$,  and the Poisson structure
 is the Poisson-Lie one\footnote{$P_{12}$ is the permutation operator (\ref{q906}).}
 \beq\label{q314}
 \begin{array}{c}
  \displaystyle{
\{S_1,S_2\}=[S_1,P_{12}]\,,
 }
 \end{array}
 \eq
coming from the classical $r$-matrix structure
 \beq\label{q315}
 \begin{array}{c}
  \displaystyle{
 \{L_1^{\rm top}(z),L_2^{\rm top}(w)\}=[L_1^{\rm top}(z)+L_2^{\rm top}(w),r_{12}(z-w)]\,,
 }
 \end{array}
 \eq
where $r_{12}(z-w)$ is the Belavin-Drinfeld $r$-matrix (see
(\ref{q102})).

The model (\ref{q310})-(\ref{q313}) possesses the  relativistic
extension \cite{LOZ8} described by equations of motion
 \beq\label{q316}
 \begin{array}{c}
  \displaystyle{
\dot S=[S,J^\eta(S)]\,,
 }
 \end{array}
 \eq
 \beq\label{q317}
 \begin{array}{c}
  \displaystyle{
 J^\eta(S)=\sum\limits_{\al\neq 0}T_\al S_\al J_\al^\eta\,,\quad
 J_\al^\eta=E_1(\eta+\om_\al)-E_1(\om_\al)
 }
 \end{array}
 \eq
and the Lax pair
  \beq\label{q318}
 \begin{array}{c}
  \displaystyle{
{L^\eta}(z)=\sum\limits_{\al} T_\al S_\al
\vf_\al(z,\om_\al+\eta)\,,\quad M^\eta(z)=-\sum\limits_{\al\neq
0}T_\al S_\al\vf_\al(z,\om_\al)\,.
 }
 \end{array}
 \eq
The Hamiltonian appears from $\tr{L^\eta}(z)$ as
 \beq\label{q319}
 \begin{array}{c}
  \displaystyle{
H^{\rm rel}=S_0\,,
 }
 \end{array}
 \eq
and the Poisson structure is the ${\rm GL}_N$ generalization of the
classical Sklyanin algebra \cite{Skl1}. It comes from the quadratic
$r$-matrix structure
 \beq\label{q320}
 \begin{array}{c}
  \displaystyle{
 \{L_1^{\eta}(z),L_2^{\eta}(w)\}=[L_1^{\eta}(z)L_2^{\eta}(w),r_{12}(z-w)]
 }
 \end{array}
 \eq
with the same $r$-matrix as in (\ref{q315}). The general (including
not only elliptic case) form of the Poisson structure follows from
the local expansion of (\ref{q320}) near $z=0$ and $w=0$, see
\cite{LOZ8}.

In case when $N-1$ eigenvalues of the matrix $S$ equal to each other
the relativistic top is gauge equivalent to the
Ruijsenaars-Schneider model (\ref{q003}), and the non-relativistic
top is gauge equivalent to the Calogero-Moser model (\ref{q007}).


\subsection{Factorization formulae}
\subsubsection*{Elliptic Ruijsenaars-Schneider model}
%
%
 The Lax matrix (\ref{q022}) is factorized as
follows
  \beq\label{q51}
  \begin{array}{l}
  \displaystyle{
 L^{\rm RS}_{ij}=\frac{\vth'(0)}{\vth(\hbar)}\sum\limits_k
 g^{-1}_{ik}(z,q)g_{kj}(z+N\hbar,q)\,e^{p_j/c}\,,
 }
 \end{array}
 \eq
where
  \beq\label{q52}
  \begin{array}{l}
  \displaystyle{
 g(z,q)=\Xi(z,q)\left(D^{0}\right)^{-1}
 }
 \end{array}
 \eq
 with
 \beq\label{q53}
 \begin{array}{c}
  \displaystyle{
\Xi_{ij}(z,q)=
 \vth\left[  \begin{array}{c}
 \frac12-\frac{i}{N} \\ \frac N2
 \end{array} \right] \left(z-Nq_j+\sum\limits_{m=1}^N
 q_m\left.\right|N\tau\right)\,,
 }
 \end{array}
 \eq
and
 \beq\label{q54}
 \begin{array}{c}
  \displaystyle{
D^0_{ij}(z,q)=\delta_{ij}D^0_{j}=\delta_{ij}
 {\prod\limits_{k\neq j}\vth(q_j-q_k)}\,.
 }
 \end{array}
 \eq
See (\ref{q907})-(\ref{q910}) for the definitions of theta-functions
with characteristics. The matrix (\ref{q53}) was introduced in
\cite{Jimbo1,Jimbo2,Jimbo3} as the intertwining matrix entering the
IRF-Vertex relations (which we review below).

Consider also the Lax matrix
  \beq\label{q55}
  \begin{array}{l}
  \displaystyle{
 {L^{\rm RS}}'_{ij}=\phi(z,q_{ij}+\hbar)\,\frac{D_j^{\hbar}}{D^0_j}\,e^{p_j/c}\,,\quad\quad
 D_j^{\hbar}=\prod_{k\neq j}{\vth(q_{jk}+\hbar)}\,,
 }
 \end{array}
 \eq
which differs from the one (\ref{q022}) by the sign of $\hbar$ in
$D^\hbar$. The Lax matrices  (\ref{q55}) and (\ref{q022}) are
related by the canonical map (\ref{q026}) with $c_1=c$ and
$c_2=\hbar$. The one (\ref{q55}) is also factorized but in a
slightly different way:
  \beq\label{q56}
  \begin{array}{l}
  \displaystyle{
 {L^{\rm RS}}'_{ij}=\frac{\vth'(0)}{\vth(\hbar)}\sum\limits_k
 \left(D^{\hbar}_i\right)^{-1}\Xi^T_{ik}(z+N\hbar,-q)\left(\Xi^T\right)_{kj}^{-1}(z,-q)D^{\hbar}_j
 \,e^{p_j/c}\,.
 }
 \end{array}
 \eq
The latter follows from (\ref{q022}) by the transposition (denoted
by $T$) and changing $q\rightarrow-q$. Curiously, both factorization
(for ${L^{\rm RS}}$ and ${L^{\rm RS}}'$) emerge in the framework of
the quantum-classical correspondence \cite{GZZ1,GZZ2,GZZ3}. They
emerge for two possible values of the $\mZ_2$-grading parameter in
the supersymmetric spin chains.

\subsubsection*{Elliptic Calogero-Moser model}

The non-relativistic limit to the Calogero-Moser model is achieved
by setting $\hbar=\nu/c$ and  $c\rightarrow\infty$ in (\ref{q51}).
This yields
  \beq\label{q0297}
  \begin{array}{c}
  \displaystyle{
 L^{\rm CM}=P+N\nu g^{-1}(z)g'(z)\,,
 }
 \end{array}
 \eq
where the non-trivial part can be written explicitly:
  \beq\label{q9501}
  \begin{array}{l}
  \displaystyle{
 \left(g^{-1}(z)g'(z)\right)_{ij}=\frac{1}{N}\,\delta_{ij}\left(E_1(z)-\sum\limits_{k\neq i}E_1(q_{ik})\right)
 +\frac{1}{N}\,(1-\delta_{ij})\phi(z,q_{ij})\,.
 }
 \end{array}
 \eq

   \subsubsection*{Trigonometric Ruijsenaars-Schneider model}

   {The Lax matrix for the trigonometric ${\rm gl}_{N}$ Ruijsenaars-Schneider model with
   spectral parameter } is of the form:
   \begin{equation}
   \label{q421}
     \displaystyle{
   L^{\rm RS}(z)_{ij}  = e^{\hbar (N-2)}\sinh(\hbar) (\coth(q_{i}-q_{j}+\hbar) + \coth(Nz))e^{p_{j}/c}
   \prod \limits_{k \neq
   j}^{N}\frac{\sinh(q_{j}-q_{k}-\hbar)}{\sinh(q_{j}-q_{k})}\,.
   }
   \end{equation}
   It admits the following factorization formula:
   \begin{equation}
   \label{q422}
     \displaystyle{
   L^{\rm RS}(z) = D^{0}\tilde{\Xi}^{-1}(z)\tilde{\Xi}(z+\hbar) (D^{0})^{-1}
   e^{P/c}\,,
   }
   \end{equation}
where
   \begin{equation}
   \label{q423}
   \begin{array}{c}
     \displaystyle{
   D^{0}_{ij} = \delta_{ij}\prod\limits_{k \neq i}(e^{-2q_{i}}-e^{-2q_{k}})\,,
 }
   \\ \ \\
     \displaystyle{
   \tilde{\Xi}_{ij}(z) = \left\{\begin{array}{l}
   x_{j}^{i-1}, \: i \leq N,\\
   \displaystyle{ x_{j}^{N-1}+\frac{(-1)^{N}}{x_{j}}, \: i = N}
   \end{array}\right.
   }
   \end{array}
   \end{equation}
   with $x_{j} = e^{-2q_{j}+2z+2\bar{q}}$. Here $\bar{q} =
   \frac{1}{N}\sum\limits_{k=1}^{N}q_{k}$ is the center of mass
   coordinate.

  {The Lax matrix for the trigonometric ${\rm gl}_{N}$ Ruijsenaars-Schneider model
  without
   spectral parameter } is of the form:
   \begin{equation}
   \label{q424}
   L^{\rm RS}_{ij} =  \frac{\sinh{\hbar}}{\sinh{(q_{i}-q_{j}+\hbar)}}e^{p_{j}/c}
   \prod\limits_{k \neq
   j}^{N}\frac{\sinh{(q_{j}-q_{k}-\hbar)}}{\sinh{(q_{j}-q_{k})}}\,.
   \end{equation}
   The factorization is as follows:
   \begin{equation}
   \label{q425}
      \begin{array}{c}
     \displaystyle{
   L^{\rm RS} =\tilde{D}^{0}({q}) \tilde{V}^{-1}({q},z) \tilde{V}({q},z+\hbar) (\tilde{D}^{0})^{-1}({q}) e^{P/c}  =
   }
   \\ \ \\
        \displaystyle{
   =\tilde{D}^{0}({q}) \tilde{V}^{-1}({q},z) Y(\hbar) \tilde{V}({q},z) (\tilde{D}^{0})^{-1}({q})
   e^{P/c}\,,
   }
   \end{array}
   \end{equation}
   where
   \begin{equation}
   \label{q4261}
   \begin{array}{l}
   \tilde{V}_{ij}(z) =\exp{((2i-1-N)(z-q_{j}))}\,,
   \end{array}
   \end{equation}
 \begin{equation}
   \label{q4262}
   \begin{array}{l}
   (\tilde{D^{0}})_{ij} = \delta_{ij}\prod\limits_{k \neq i}\sinh(q_{i}-q_{k})
   \end{array}
   \end{equation}
and
 \begin{equation}
   \label{q4263}
   \begin{array}{l}
   Y(\lambda)_{ij} =\delta_{ij} \exp{(-(N+1-2i)\lambda)}\,.
   \end{array}
   \end{equation}

   \subsubsection*{Trigonometric Calogero-Moser model}
   The {Lax matrix of the trigonometric ${\rm gl}_{N}$ Calogero-Moser model with spectral parameter} is of the following form:
   \begin{equation}
   \label{q427}
      \begin{array}{c}
     \displaystyle{
   L^{\rm CM}_{ij}(z)=\delta_{ij}\left(p_{i} + \nu (N-2) + \nu \coth(Nz)-\nu \sum\limits_{k \neq i}^{N}\coth(q_{i}-q_{k})\right) +
   }
   \\ \ \\
        \displaystyle{
+   \nu(1-\delta_{ij}) (\coth(q_{i}-q_{j}) + \coth(Nz))\,.
    }
   \end{array}
   \end{equation}
   The factorization formula is as follows:
   \begin{equation}
   \label{q428}
   L^{\rm CM}(z) = P+\nu D^{0} \tilde{\Xi}^{-1}(\partial_{z}\tilde{\Xi})
   (D^{0})^{-1}\,,
   \end{equation}
where $\tilde{\Xi}$ and $D^{0}$ are defined in (\ref{q423}).

   The {Lax matrix of the trigonometric ${\rm gl}_{N}$ Calogero-Moser model without spectral parameter} is of the following form:
   \begin{equation}
   \label{q429}
   L^{\rm CM} = \delta_{ij}(p_{i}-\nu \sum\limits_{k \neq i}^{N}\coth{(q_{i}-q_{k})})
   +(1-\delta_{ij})\frac{\nu}{\sinh{(q_{i}-q_{j})}}\,.
   \end{equation}
   Its factorization is as follows:
   \begin{equation}
   \label{q430}
   L^{\rm CM} = P + \nu \tilde{D}^{0}\tilde{V}^{-1}(\lambda)\partial_{\lambda}\tilde{V}(\lambda)(\tilde{D}^{0})^{-1} =
   P +
   \tilde{D}^{0}\tilde{V}^{-1}(\log Y)\tilde{V}(\tilde{D}^{0})^{-1},
   \end{equation}
   where
   \beq\label{q4301}
     (\log Y)_{ij} = \delta_{ij}\, \nu (2i-1-N)
   \eq
while $\tilde{V}$ and  $\tilde{D}^{0}$ are those from (\ref{q4261})
and (\ref{q4262}).

   \subsubsection*{ Rational Ruijsenaars-Schneider model }
   {The Lax matrix for the rational ${\rm gl}_{N}$ Ruijsenaars-Schneider model
  with
   spectral parameter } is of the form:
   \begin{equation}
   \label{q431}
        \displaystyle{
   L_{ij}^{\rm RS}(z) = \hbar\, \left(\frac{1}{q_{i}-q_{j}+\hbar} + \frac{1}{Nz}\right) e^{p_{j}/c}
   \prod\limits_{k \neq
   j}^{N}\frac{q_{j}-q_{k}-\hbar}{q_{j}-q_{k}}\,.
   }
   \end{equation}
   It admits the following factorization formula:
   \begin{equation}
   \label{q432}
   L^{\rm RS}(z) = D^{0}({q}) \Xi^{-1}({q},z) \Xi({q},z+\hbar) (D^{0})^{-1}({q})
   e^{P/c}\,,
   \end{equation}
   where
   \begin{equation}
   \label{q433}
   \begin{array}{c}
   (D^{0})_{ij}({q}) =\delta_{ij}\prod\limits_{k \neq i}^{n}(q_{i}-q_{k})\,,
     \\
   \Xi_{ij}({q},z) =(z-q_{j}+\bar{q})^{\varrho(i)}\,,\quad \bar{q} = \frac{1}{N}\sum\limits_{k=1}^{N}q_{k}
   \end{array}
   \end{equation}
   with
   \begin{equation}
   \label{q434}
   \varrho(i) = \left\{\begin{array}{l}
   i-1 \;  \hbox{for} \; 1 \leq i \leq N-1,\\
   N\;  \hbox{for} \; i=N\,.
   \end{array}\right.
   \end{equation}

      {The Lax matrix for the rational ${\rm gl}_{N}$ Ruijsenaars-Schneider model
  without
   spectral parameter } is of the form:
   \begin{equation}
   \label{q435}
   L^{\rm RS}_{ij} =  \frac{\hbar\, e^{p_{j}/c}}{q_{i}-q_{j}+\hbar}  \prod\limits_{k \neq
   j}^{N}\frac{q_{j}-q_{k}-\hbar}{q_{j}-q_{k}}\,.
   \end{equation}
   It admits the following factorization formula:
   \begin{equation}
   \label{q436}
   \begin{array}{c}
           \displaystyle{
   L^{\rm RS} =D^{0}({q}) V^{-1}(z) V(z+\hbar) (D^{0})^{-1}({q}) e^{P/c}  =
    }
    \\ \ \\
            \displaystyle{
   =D^{0}({q}) V^{-1}(z) C_{\hbar} V(z) (D^{0})^{-1}({q}) e^{P/c}  ,
    }
    \end{array}
   \end{equation}
   where
   \begin{equation}
   \label{q4371}
   \begin{array}{c}
   V_{ij}(z) =(z-q_{j}+\bar{q})^{i-1}
   \end{array}
   \end{equation}
   is the Vandermonde matrix, $D^{0}({q})$ is (\ref{q433})  and
   \begin{equation}
   \label{q4372}
   \begin{array}{c}
   \left(C_{\lambda}\right)_{ij} =\left\{\begin{array}{l}
   \displaystyle{\frac{(i-1)!\lambda^{i-j}}{(j-1)!(i-j)!}, \;  j \leq i}\,,\\
   0 , \;  j > i\,.
   \end{array}\right.
   \end{array}
   \end{equation}
   The following simplification of (\ref{q436}) is also correct:
   \begin{equation}
   \label{q438}
   L^{\rm RS} =
   D^{0}({q})V^{-1}({q})C_{\hbar}V({q})(D^{0})^{-1}({q})e^{P/c}\,,
   \end{equation}
   where
   \begin{equation}
   \label{q439}
   V_{ij}({q}) = (-q_{j})^{i-1}\,.
   \end{equation}

   \subsubsection*{Rational Calogero-Moser Model}
   The {Lax matrix of the rational ${\rm gl}_{N}$ Calogero-Moser model with spectral parameter} is of the following form:
   \begin{equation}
   \label{q440}
   L^{\rm CM}_{ij}(z) = \delta_{ij}(p_{i}-\sum\limits_{k \neq i}^{N}\frac{\nu}{q_{i}-q_{k}})+
   (1-\delta_{ij})\frac{\nu}{q_{i}-q_{j}}+\frac{\nu}{Nz}\,.
   \end{equation}
The factorization formula is as follows:
   \begin{equation}
   \label{q441}
   L^{\rm CM}(z) = P +\nu
   D^{0}\Xi^{-1}(\partial_{z}\Xi)(D^{0})^{-1}\,,
   \end{equation}
where $\Xi$ and $D^{0}$ are those from (\ref{q433}).

   The {Lax matrix of the rational ${\rm gl}_{N}$ Calogero-Moser model without spectral parameter} is of the following form:
   \begin{equation}
   \label{q443}
   L^{\rm CM}_{ij} = \delta_{ij}(p_{i}-\sum\limits_{k \neq
   i}^{N}\frac{\nu}{q_{i}-q_{k}})+(1-\delta_{ij})\frac{\nu}{q_{i}-q_{j}}\,.
   \end{equation}
Its factorization is given by
   \begin{equation}
   \label{q444}
   L^{\rm CM}(z) = P +\nu D^{0}V^{-1}(z)(\partial_{z}V)(z)(D^{0})^{-1}\,,
   \end{equation}
   where $D^{0}$ is defined in (\ref{q433}) and $V(z)$ -- in
   (\ref{q4371}).
Equivalently, one can represent (\ref{q443}) in the form:
   \begin{equation}
   \label{q446}
   L^{\rm CM} = P+\nu D^{0}V^{-1}({q}) C_{0} V({q})(D^{0})^{-1},
   \end{equation}
   with $V({q})$ (\ref{q439})
   and
     \begin{equation}
   \label{q447}
   \begin{array}{c}
   (C_{0})_{ij}=
   \left\{\begin{array}{l}
   i-1,  i=j+1\,,\\
   0,  \hbox{otherwise}\,.
   \end{array}\right.
   \end{array}
   \end{equation}

\section{IRF-Vertex relations}\label{sect3}
\setcounter{equation}{0}
 %
\subsection{IRF-Vertex correspondence}
First, let us introduce three quantum $R$-matrices.

\noindent{\bf Baxter-Belavin (non-dynamical) $R$-matrix}
\cite{Baxter,Belavin} (see also \cite{RT}):
 \beq\label{q100}
 \begin{array}{c}
  \displaystyle{
R_{12}^\hbar(z)=\sum\limits_\al T_\al\otimes T_{-\al}\,
\vf_\al(z,\om_\al+\hbar)\,,\quad\quad
\res\limits_{z=0}R_{12}^\hbar(z)=NP_{12}\,.
  }
 \end{array}
 \eq
The classical limit (near $\hbar=0$)
 \beq\label{q101}
 \begin{array}{c}
  \displaystyle{
R_{12}^\hbar(z)=\frac{1\otimes 1}{\hbar}+r_{12}(z)+\hbar\, m_{12}(z)
+O(\hbar^2)
  }
 \end{array}
 \eq
provides the classical Belavin-Drinfeld \cite{BD} $r$-matrix
 \beq\label{q102}
 \begin{array}{c}
  \displaystyle{
r_{12}(z)=1\otimes 1 E_1(z)+\sum\limits_{\al\neq 0} T_\al\otimes
T_{-\al}\, \vf_\al(z,\om_\al)\,.
  }
 \end{array}
 \eq
The Baxter-Belavin $R$-matrix
$R^\hbar_{12}(z_1,z_2)=R^\hbar_{12}(z_1-z_2)$ (\ref{q100}) satisfies
the quantum Yang-Baxter equation
  \beq\label{q103}
  \begin{array}{c}
  \displaystyle{
 R^\hbar_{12}(z_1,z_2)R^\hbar_{13}(z_1,z_3)R^\hbar_{23}(z_2,z_3)
 =R^\hbar_{23}(z_2,z_3)R^\hbar_{13}(z_1,z_3)R^\hbar_{12}(z_1,z_2)
 }
 \end{array}
 \eq
In this Section we will also use notation
 \beq\label{q104}
 \begin{array}{c}
  \displaystyle{
 R^{\hbox{\tiny{B}}}_{12}(\hbar,z_1,z_2)= R^{\hbox{\tiny{B}}}_{12}(\hbar,z_1-z_2)=
 \frac{1}{N}\,R_{12}^{\hbar/N}(z_1-z_2)\,.
 }
 \end{array}
 \eq
%
{\bf Felder's (dynamical) $R$-matrix} \cite{Felder2}:
 \beq\label{q105}
 \begin{array}{c}
  \displaystyle{
 R^{\hbox{\tiny{F}}}_{12}(\hbar,z_1,z_2|\,q)=R^{\hbox{\tiny{F}}}_{12}(\hbar,z_1-z_2|\,q)=
 }
\\ \ \\
  \displaystyle{
 =\sum\limits_{i\neq j}
 E_{ii}\otimes E_{jj}\, \phi(\hbar,-q_{ij})+\sum\limits_{i\neq j}
 E_{ij}\otimes E_{ji}\, \phi(z_1-z_2,q_{ij})+\phi(\hbar,z_1-z_2)\sum\limits_{i}
 E_{ii}\otimes E_{ii}\,.
 }
 \end{array}
 \eq
 It is a solution of the quantum dynamical Yang-Baxter equation
  \beq\label{q106}
  \begin{array}{c}
  \displaystyle{
 R^\hbar_{12}(z_1,z_2|\,q)R^\hbar_{13}(z_1,z_3|\,q-\hbar^{(2)})R^\hbar_{23}(z_2,z_3|\,q)=\hspace{40mm}
}
\\ \ \\
  \displaystyle{
\hspace{40mm}
=R^\hbar_{23}(z_2,z_3|\,q-\hbar^{(1)})R^\hbar_{13}(z_1,z_3|\,q)R^\hbar_{12}(z_1,z_2|\,q-\hbar^{(3)})\,,
 }
 \end{array}
 \eq
where the shifts of dynamical arguments $u$ are performed as
follows:
  \beq\label{q107}
  \begin{array}{c}
  \displaystyle{
R^\hbar_{12}(z_1,z_2|\,q+\hbar^{(3)})=P_3^\hbar\,
R^\hbar_{12}(z_1,z_2|\,q)\, P_3^{-\hbar} \,,\quad
P_3^\hbar=\sum\limits_{k=1}^N 1\otimes 1\otimes E_{kk}
\exp(\hbar\frac{\p}{\p q_k})\,.
 }
 \end{array}
 \eq
%
\noindent {\bf Arutyunov-Chekhov-Frolov (semi-dynamical) $R$-matrix}
\cite{Arut}:
 \beq\label{q108}
 \begin{array}{c}
  \displaystyle{
 R^{\hbox{\tiny{ACF}}}_{12}(\hbar,z_1,z_2|\,q)=\sum\limits_{i\neq j}
 E_{ii}\otimes E_{jj}\, \phi(\hbar,q_{ij})+\sum\limits_{i\neq j}
 E_{ij}\otimes E_{ji}\, \phi(z_1-z_2,q_{ij})-
 }
\\ \ \\
  \displaystyle{
-\sum\limits_{i\neq j}
 E_{ij}\otimes E_{jj}\, \phi(z_1+\hbar,q_{ij})+\sum\limits_{i\neq j}
 E_{jj}\otimes E_{ij}\, \phi(z_2,q_{ij})+
 }
\\ \ \\
  \displaystyle{
+(E_1(\hbar)+E_1(z_1-z_2)+E_1(z_2)-E_1(z_1+\hbar))\sum\limits_{i}
 E_{ii}\otimes E_{ii}
 }
 \end{array}
 \eq
satisfies the following (semi-dynamical) Yang-Baxter equation:
  \beq\label{q109}
  \begin{array}{c}
  \displaystyle{
 R^\hbar_{12}(z_1,z_2|\,q)R^\hbar_{13}(z_1-\hbar,z_3-\hbar|\,q)R^\hbar_{23}(z_2,z_3|\,q)=\hspace{50mm}
}
\\ \ \\
  \displaystyle{
\hspace{40mm}
=R^\hbar_{23}(z_2-\hbar,z_3-\hbar|\,q)R^\hbar_{13}(z_1,z_3|\,q)R^\hbar_{12}(z_1-\hbar,z_2-\hbar|\,q)\,.
 }
 \end{array}
 \eq

\noindent {\bf IRF-Vertex correspondence}
\cite{Baxter2,Jimbo1,Jimbo2,Jimbo3,Hasegawa1,Hasegawa2,Hasegawa3}
establishes an explicit relationship between dynamical and
non-dynamical $R$-matrices (\ref{q104}) and (\ref{q105}):
 \beq\label{q110}
 \begin{array}{c}
  \displaystyle{
g_2(z_2,q)\,
g_1(z_1,q+\hbar^{(2)})\,R^{\hbox{\tiny{F}}}_{12}(\hbar,z_1-z_2|\,q)=R^{\hbox{\tiny{B}}}_{12}(\hbar,z_1-z_2)\,
g_1(z_1,q)\, g_2(z_2,q+\hbar^{(1)})\,.
 }
 \end{array}
 \eq
For the semi-dynamical $R$-matrix (\ref{q108}) the following
relations hold true \cite{Arut,SeZ}:
  \beq\label{q111}
  \begin{array}{c}
  \displaystyle{
 R^{\hbox{\tiny{F}}}_{12}(\hbar,z_1-z_2|\,q)=
 }
 \\ \ \\
  \displaystyle{
 =g_1^{-1}(z_1,q+\hbar^{(2)})\,
 g_1(z_1+\hbar,q)\,R^{\hbox{\tiny{ACF}}}_{12}(\hbar,z_1,z_2|\,q)\,
 g_2^{-1}(z_2+\hbar,q)\, g_2(z_2,q+\hbar^{(1)})\,.
 }
 \end{array}
 \eq
Combining (\ref{q110}) and (\ref{q111}) we get
  \beq\label{q112}
  \begin{array}{c}
  \displaystyle{
 R^{\hbox{\tiny{B}}}_{12}(\hbar,z_1-z_2)=g_1(z_1+\hbar,q)\,
 g_2(z_2,q)\,
 R^{\hbox{\tiny{ACF}}}_{12}(\hbar,z_1,z_2|\,q)\, g_2^{-1}(z_2+\hbar,q)\,
 g_1^{-1}(z_1,q)\,.
 }
 \end{array}
 \eq
Following \cite{SeZ} let us rewrite relation (\ref{q112}) as
  \beq\label{q113}
  \begin{array}{c}
  \displaystyle{
 g_2^{-1}(z_2,q)\,R^{\hbox{\tiny{B}}}_{12}(\hbar,z_1-z_2)=g_1(z_1+\hbar,q)\,
 R^{\hbox{\tiny{ACF}}}_{12}(\hbar,z_1,z_2|\,q)\, g_2^{-1}(z_2+\hbar,q)\,
 g_1^{-1}(z_1,q)
 }
 \end{array}
 \eq
and take the residues at $z_2=0$ of both parts:
  \beq\label{q114}
  \begin{array}{c}
  \displaystyle{
 {\breve g}_2(0,q)\,R^{\hbox{\tiny{B}}}_{12}(\hbar,z)=g_1(z+\hbar,q)\,
 \mathcal O_{12}\, g_2^{-1}(\hbar,q)\,
 g_1^{-1}(z,q)\,,
 }
 \end{array}
 \eq
 where
   \beq\label{q115}
  \begin{array}{c}
  \displaystyle{
{\breve g}(0,q)=\res\limits_{z=0}\,g^{-1}(z)
 }
 \end{array}
 \eq
and
   \beq\label{q116}
  \begin{array}{c}
  \displaystyle{
 \mathcal O_{12}=\res\limits_{z_2=0}\,R^{\hbox{\tiny{ACF}}}_{12}(\hbar,z_1,z_2|\,q)=\sum\limits_{i,j}E_{ii}\otimes
E_{ji}\,.
 }
 \end{array}
 \eq
Then, for the $R$-matrix (\ref{q100}) we have\footnote{Notice that
for $N=1$ (\ref{q117}) reproduces the definition of the Kronecker
function (\ref{q911}).}
  \beq\label{q117}
  \begin{array}{c}
  \displaystyle{
 \frac1N\,{\breve g}_2(0,q)\,R^\hbar_{12}(z)=g_1(z+N\hbar,q)\,
 \mathcal O_{12}\, g_2^{-1}(N\hbar,q)\,
 g_1^{-1}(z,q)\,.
 }
 \end{array}
 \eq
This formula is $R$-matrix analogue of the Lax matrix factorization.
We will use it in Section \ref{Sect35} for evaluation of the
$M$-matrix.

\subsection{Classical IRF-Vertex relations}
%
{The classical IRF-Vertex transformations} relate the classical
dynamical $r$-matrix structures of the Ruijsenaars-Schneider (or
Calogero-Moser) models with the non-dynamical $r$-matrix structures
of the relativistic top (\ref{q320}) (or the non-relativistic top
(\ref{q315})), see e.g.
\cite{BFeher1,BFeher2,BFeher3,Chen11,Chen12}. At the level of the
classical Lax matrices the IRF-Vertex transformation is the gauge
transformation generated by the matrix $g(z)$:
 \beq\label{q363}
 \begin{array}{c}
  \displaystyle{
 L^{\rm top}(z)=g(z)L^{\rm CM}(z)g^{-1}(z)\,,
 }
 \end{array}
 \eq
for the Lax matrices (\ref{q029}) and (\ref{q312}). Similarly,
 \beq\label{q364}
 \begin{array}{c}
  \displaystyle{
 L^{\hbar}(z)=g(z)L^{\rm RS}(z)g^{-1}(z)\,,
 }
 \end{array}
 \eq
for the Lax matrices (\ref{q51}) and (\ref{q317}). Being written as
(\ref{q363}) and (\ref{q364}) these tops are just alternative forms
of the Ruijsenaars-Schneider and Calogero-Moser models respectively.
However, these are only special cases of the tops corresponding to
the rank one matrix $S$. In the general case the dimensions of the
phase spaces of the tops are large than those for the spinless
many-body systems.

\paragraph{Structure of the intertwining matrix.}
%
The intertwining matrix $g(z)$ (\ref{q52}) satisfies the following
properties
\cite{Hasegawa1,Hasegawa2,Hasegawa3,Jimbo1,Jimbo2,Jimbo3,RT}:

1. The matrix $g(z)$ is degenerated at $z=0$. See (\ref{q952}).

2. The matrix $g(0)$ has one-dimensional kernel in the direction of
the vector-column
 \beq\label{q360}
 \begin{array}{c}
  \displaystyle{
 \rho=(1,1,...,1)^T\in\mC^N
 }
 \end{array}
 \eq

Consider $g^{-1}(z)$ near $z=0$:
 \beq\label{q361}
 \begin{array}{c}
  \displaystyle{
 g^{-1}(z)=\frac1z\,{\breve g}(0)+A+O(z)\,,\quad {\breve
 g}(0,q)=\res\limits_{z=0}\,g^{-1}(z)\,.
 }
 \end{array}
 \eq
Then the matrix ${\breve g}(0)$ is of rank one:
 \beq\label{q362}
 \begin{array}{c}
  \displaystyle{
 {\breve g}(0)=\rho\otimes\psi\,,\quad \psi=(\psi_1(q),...,\psi_N(q))\in\mC^N\,.
 }
 \end{array}
 \eq
 and
 \beq\label{q3621}
 \begin{array}{c}
  \displaystyle{
 \psi=\frac1N\,\rho^T{\breve g}(0)\,.
 }
 \end{array}
 \eq
 Indeed, by expanding $g^{-1}(z)g(z)=1_N$ near $z=0$ we get ${\breve
 g}(0)g(0)=0$. The kernel of  $g(0)$ is one-dimensional. Therefore,
 the kernel of ${\breve  g}(0)$ is $N-1$ dimensional. The latter
 means that ${\breve  g}(0)$ is a product of a vector by covector.
 On the other hand, $g(0){\breve
 g}(0)=0$. Thus, the vector should lie in the kernel $g(0)$, i.e. it is proportional to $\rho$ (\ref{q360}).
 This gives (\ref{q362}).
%
%
%

\paragraph{Classical bosonization formulae} are the classical analogues of the
representation of the Sklyanin algebra generators in terms of the
difference operators, i.e. the top's variables $S_\al$ (entering
${\rm GL}_N$ classical Sklyanin algebra) are expressed in terms of
the Ruijsenaars-Schneider variables. The non-relativistic limit
leads to the classical Lie (co)algebra variables expressed in terms
of the Calogero-Moser variables. For the explicit change of
variables see \cite{Skl2,Hasegawa1,Hasegawa2,Chen11,Chen12} and
\cite{AASZ,LOZ8,KrZ}.

In the above formulae (\ref{q363}), (\ref{q364}) the top models are
of very special type. The matrices $S$ in both cases are of rank
one, while in (\ref{q312}) and/or (\ref{q317}) they are arbitrary.
Indeed, the matrices $S$ are residues of the corresponding Lax
matrices. Assuming (\ref{q51})-(\ref{q52})
 \beq\label{q365}
 \begin{array}{c}
  \displaystyle{
 L^{\rm RS}=\frac{\vth'(0)}{\vth(\hbar)}\,
 g^{-1}(z,q)g(z+N\hbar,q)\,e^{P/c}\,,\quad\quad P={\rm diag}(p_1,...,p_N)
 }
 \end{array}
 \eq
and (\ref{q364}) we get
 \beq\label{q366}
 \begin{array}{c}
  \displaystyle{
 L^{\hbar}(z)=\frac{\vth'(0)}{\vth(\hbar)}\,
 g(z+N\hbar,q)\,e^{P/c}g^{-1}(z,q)\,.
 }
 \end{array}
 \eq
Therefore, for the matrix $S$ of dynamical variables in the
relativistic top we have
 \beq\label{q367}
 \begin{array}{c}
  \displaystyle{
 S=\res\limits_{z=0}L^{\hbar}(z)=\frac{\vth'(0)}{\vth(\hbar)}\,
 g(N\hbar)\,e^{P/c}\,{\breve g}(0)\stackrel{(\ref{q362})}{=}\xi\otimes \psi\,,
 }
 \end{array}
 \eq
where
 \beq\label{q368}
 \begin{array}{c}
  \displaystyle{
 \xi=\frac{\vth'(0)}{\vth(\hbar)}\,
 g(N\hbar)\,e^{P/c}\,\rho\,,\quad\quad \psi=\frac1N\,\rho^T{\breve g}(0)\,.
 }
 \end{array}
 \eq
The point of the phenomenon is that the Lax matrix (\ref{q366}) is
expressed through the variables $S$ (\ref{q367}).

The row-vector $\psi$ can be found in a
different way. The residue of the Ruijsenaars-Schneider Lax matrix
(\ref{q022}) is of the form:
 \beq\label{q369}
 \begin{array}{c}
  \displaystyle{
 \res\limits_{z=0}L^{\rm RS}(z)=\rho\otimes\rho^T
 D^{-\hbar}(D^0)^{-1}e^{P/c}\,.
 }
 \end{array}
 \eq
On the other hand, from (\ref{q365}) we have
 \beq\label{q370}
 \begin{array}{c}
  \displaystyle{
 \res\limits_{z=0}L^{\rm RS}(z)=\frac{\vth'(0)}{\vth(\hbar)}\,
 {\breve g}(0)g(N\hbar)\,e^{P/c}=\frac{\vth'(0)}{\vth(\hbar)}\,
 \rho\otimes\psi g(N\hbar)\,e^{P/c}\,.
 }
 \end{array}
 \eq
By comparing (\ref{q369}) and (\ref{q370}) we come to
 \beq\label{q371}
 \begin{array}{c}
  \displaystyle{
 \psi=\frac{\vth'(0)}{\vth(\hbar)}\,\rho^T
 D^{-\hbar}(D^0)^{-1} g^{-1}(N\hbar)\,.
 }
 \end{array}
 \eq
Notice that by the definition (\ref{q362}) $\psi$ is independent of
$\hbar$. Therefore, the parameter $\hbar$ in the r.h.s. of
(\ref{q371}) is arbitrary. Tending it to zero we reproduce
(\ref{q3621}). Plugging (\ref{q371}) into (\ref{q367}) we get
 \beq\label{q372}
 \begin{array}{c}
  \displaystyle{
 S=g(N\hbar)\,e^{P/c}\,\rho\otimes\rho^T D^{-\hbar}(D^0)^{-1}
 g^{-1}(N\hbar)\,.
 }
 \end{array}
 \eq
 In the non-relativistic limit the generators of the Poisson-Lie
 structure appear. By setting $\hbar=\nu/c$ and taking the limit   $c\rightarrow\infty$
 in (\ref{q367}) we obtain:
 \beq\label{q3721}
 \begin{array}{c}
  \displaystyle{
 S=g(0)P{\breve g}(0)+N\nu g'(0){\breve g}(0)=\mu\otimes\psi\,,\quad\quad \mu=(g(0)P+N\nu g'(0))\rho\,.
 }
 \end{array}
 \eq
The Poisson-Lie brackets for $S$ follow from the canonical brackets
between components of $\mu$ and  $\psi$:
$\{\mu_i,\psi_j\}\propto\delta_{ij}$. This lead to a natural
quantization $\hat{\mu}_i\propto \p/\p\psi_i$. Such coordinates were
used in \cite{Turbiner} and \cite{Perelomov1,Perelomov2} for
reformulation of the quantum Calogero-Moser model.

\paragraph{Modifications of bundles.} The IRF-Vertex intertwining matrix can be treated as  modification of bundle
\cite{Vakul,LOZ}.
It is no coincidence that the vector $\rho$ (\ref{q360})  enters the
residue of the Lax matrix (\ref{q369}). In fact, dealing with the
singular gauge transformation (degenerated at point $z=0$) we must
impose condition for  an eigenvector ($\rho$) of the residue of the
Lax matrix under transformation to lie in the kernel of the gauge
transformation at point $z=0$: $\rho\in{\rm Ker}\,g(0)$.  This
condition
 comes from the requirement not to produce the second order pole at
 $z=0$ when performing conjugation by the matrix $g(z)$. We explain
it below. Here, for the Lax matrix (\ref{q369}) it is
 easy to see by expansion of the r.h.s. of (\ref{q364}) near $z=0$.
 The vanishing of the second order pole is equivalent to
 \beq\label{q373}
 \begin{array}{c}
  \displaystyle{
 g(0)\res\limits_{z=0}L^{\rm RS}(z)\,{\breve g}(0)=0\,.
 }
 \end{array}
 \eq
It is fulfilled due to $\res\limits_{z=0}L^{\rm
RS}(z)\rho=\lambda_0\rho$, $\lambda_0=\sum_k{\dot q}_k=H^{\rm RS}/c$
and $g(0)\rho=0$.

The Lax matrices with spectral parameter $z$ can be viewed as
sections of bundles over the base curve $\Sigma$ with local
coordinate $z$ \cite{Krich0,Krich1}. In our case $\Sigma$ is the
elliptic curve with moduli $\tau$. The Lax matrices are fixed by
their residues and quasi-periodic behavior on the lattice
$\mZ+\tau\mZ$. The latter means that they are sections of ${\rm
End}(V)$-bundles for some holomorphic vector bundles $V$. For the
Lax matrices of the Calogero-Moser (\ref{q029}) and the elliptic top
(\ref{q312}) models using (\ref{q9122}) we have
 \beq\label{q374}
 \begin{array}{c}
  \displaystyle{
 L^{\rm CM}(z+1)= L^{\rm CM}(z)\,,\quad\quad L^{\rm CM}(z+\tau)=
   e^{-2\pi\imath\, {\rm diag}(q)}L^{\rm CM}(z)e^{2\pi\imath\,  {\rm diag}(q)}\,,
 }
 \end{array}
 \eq
where
 \beq\label{q3741}
 \begin{array}{c}
  \displaystyle{
 {\rm diag}(q)={\rm diag}(q_1,q_2...,q_N)\in\Mat
 }
 \end{array}
 \eq
is the diagonal matrix built of coordinates of particles, and
 \beq\label{q375}
 \begin{array}{c}
  \displaystyle{
 L^{\rm top}(z+1)= Q^{-1}L^{\rm top}(z)Q\,,\quad\quad L^{\rm top}(z+\tau)=
  \Lambda^{-1} L^{\rm top}(z)\Lambda\,,
 }
 \end{array}
 \eq
where $Q,\Lambda$ are the matrices (\ref{q902}). In the relativistic
case an additional factor
$\exp(-2\pi\imath\hbar)$ appears for the shift of $z$ by $\tau$. It
can be removed by dividing the Lax matrix by function
$\phi(z,\hbar)$.

The IRF-Vertex transformation acts as gauge transformation, which
changes the quasi-periodic properties from (\ref{q374}) to
(\ref{q375}). In fact, this condition almost fixes the matrix $g(z)$
(\ref{q52}). More precisely, it fixes the $\Xi(z)$ part of $g(z)$,
while the $D^0$ factor comes from the discussed above requirement
for the vector $\rho$ to belong to the kernel of $g(0)$.

The map between two bundles, which is an isomorphism everywhere
except a point, where it has one-dimensional kernel is known as the
modification of (the initial) bundle \cite{Dr2,Arinkin1,Arinkin2}.
In  our case it is performed at point $z=0$ in the direction $\rho$.
Locally the modification is described as follows. Let us choose the
basis in of sections in a way that the residue $L_{-1}$ at $z=0$ of
the initial Lax matrix $L(z)\in\Gamma({\rm End}(V))$ is of the form
 \beq\label{q376}
 \begin{array}{c}
  \displaystyle{
 L_{-1}=\mat{\lambda}{{*}_{1\times(N-1)}}{0_{(N-1)\times
 1}}{*_{(N-1)\times(N-1)}}\,.
 }
 \end{array}
 \eq
Then its eigenvector is $v=(1,0,...,0)^T$: $L_{-1} v=\lambda v$. The
modification towards this direction is given by
 \beq\label{q377}
 \begin{array}{c}
  \displaystyle{
 g(z)=\mat{z}{{0}_{1\times(N-1)}}{0_{(N-1)\times
 1}}{1_{(N-1)\times(N-1)}}\,.
 }
 \end{array}
 \eq
In this case $g(0)v=0$ and $\res\limits_{z=0}g^{-1}(z)=v\otimes v^T$
-- rank one matrix. We also have
 \beq\label{q378}
 \begin{array}{c}
  \displaystyle{
 g(z) L_{-1}g^{-1}(z)=\mat{\lambda}{z{*}_{1\times(N-1)}}{0_{(N-1)\times
 1}}{*_{(N-1)\times(N-1)}}\,.
 }
 \end{array}
 \eq
This demonstrates that the second order pole does not appear. Notice
also that the transformation (\ref{q377}) adds the zero at $z=0$ to
the section of the  ${\rm det}V$. This results in changing degree of
the initial vector bundle $V$ by one. So that the Calogero-Moser
model correspond to ${\rm deg}V=0$, while the elliptic top model --
to ${\rm deg}V=1$. The vector bundles over elliptic curves were
classified in \cite{Atiyah}. In the Hitchin approach
\cite{Hitchin1,Hitchin2,Hitchin3} to elliptic integrable systems the
moduli space of the underlying vector bundles play the role of the
configuration space of the integrable system. Its dimension is equal
to g.c.d.$({\rm rk}(V),{\rm deg}(V))$. This could be understood as
follows. For the ${\rm deg}(V)=k$ bundle the quasi-periodic
properties of the Lax matrix are
 \beq\label{q379}
 \begin{array}{c}
  \displaystyle{
 L^{\rm top}(z+1)= Q^{-1}L^{\rm top}(z)Q\,,\quad\quad L^{\rm top}(z+\tau)=
  \Lambda^{-k} L^{\rm top}(z)\Lambda^k\,.
 }
 \end{array}
 \eq
If g.c.d.$(N,k)=m>1$ then there exist a matrix $X$ parameterized by
$m$ variables ($q_i$) with the property $[Q,X]=[\Lambda^k,X]=0$, so
that the boundary conditions (\ref{q379}) become degenerated. This
degeneracy can be eliminated by redefinition of (\ref{q379}) as
 \beq\label{q3791}
 \begin{array}{c}
  \displaystyle{
 L^{\rm top}(z+1)= Q^{-1}L^{\rm top}(z)Q\,,\quad\quad L^{\rm top}(z+\tau)=
  X^{-1}\Lambda^{-k} L^{\rm top}(z)\Lambda^kX\,.
 }
 \end{array}
 \eq
By reexpressing $X$ through $m$ variables of $q_i$ type we get a
model representing an intermediate case between the many-body and
the tops systems \cite{LOSZ4}. Thus, starting with $k=0$ and
increasing the degree of $V$ by modifications provides a family of
gauge equivalent integrable Hitchin type systems including the
(spin) Calogero-Moser model and the elliptic top. This scheme was
called the symplectic Hecke correspondence \cite{LOZ,SZ}, and it is
naturally generalized to the case when the structure group of the
principle bundle (associated with the vector bundle $V$) is  an
arbitrary complex simple Lie group \cite{LOSZ1,LOSZ2}.

\subsection{Factorization of the Lax matrix}
%

To proceed we need the $R$-matrix formulation for the tops models
\cite{LOZ8}\footnote{The general idea is similar to the
quasi-classical description presented originally in \cite{Skl1}.}
The Lax pair of the relativistic top model (\ref{q318}) can be
written in terms of the Belavin's $R$-matrix
(\ref{q100})-(\ref{q102}) as follows\footnote{Here the $M$-matrix
differs from the one in (\ref{q318}) by the term proportional to
identity matrix (it is cancelled out from the Lax equations).}:
  \beq\label{q201}
  \begin{array}{c}
  \displaystyle{
L^\hbar(S,z)=\frac1N\,\tr_2\left(R_{12}^\hbar(z)S_2\right)\,,\quad\quad
M^\hbar(S,z)=-\frac1N\,\tr_2\left(r_{12}(z)S_2\right)\,.
 }
 \end{array}
 \eq
 The factor $1/N$ comes from (\ref{q905}). In fact, the
formulae (\ref{q201}) are valid for a wider class of integrable
tops, which appear when the underlying $R$-matrix satisfies the
associative Yang-Baxter equation together with appropriate classical
limit and skew-symmetry and/or unitarity conditions \cite{LOZ16}.

Multiply both sides of (\ref{q117}) by
$g_2(N\hbar)e^{P_2/c}\vth'(0)/\vth(\hbar)$ from the left:
  $$
  \displaystyle{
 \frac{\vth'(0)}{N\vth(\hbar)}\,g_2(N\hbar)e^{P_2/c}{\breve g}_2(0,q)\,R^\hbar_{12}(z)=
 \frac{\vth'(0)}{\vth(\hbar)}\,g_2(N\hbar)e^{P_2/c}g_1(z+N\hbar,q)\,
 \mathcal O_{12}\, g_2^{-1}(N\hbar,q)\,
 g_1^{-1}(z,q)\,.
 }
 $$
The trace over the second space provides in the first space the Lax
matrix (\ref{q201}) with $S=S(p,q)$ (\ref{q367}):
  \beq\label{q202}
  \begin{array}{c}
  \displaystyle{
L_1^\hbar(S(p,q),z)=\tr_2\left(\frac{\vth'(0)}{\vth(\hbar)}\,e^{P_2/c}g_1(z+N\hbar,q)\,
 \mathcal O_{12}\, g_1^{-1}(z,q)\right)\,.
 }
 \end{array}
 \eq
Taking into account that for any matrix
$T=\sum_{i,j}E_{ij}T_{ij}\in\Mat$
  \beq\label{q203}
  \begin{array}{c}
  \displaystyle{
\tr_2 \left(\mathcal O_{12}T_2\right)=\sum\limits_i
E_{ii}\sum\limits_j T_{ij}
 }
 \end{array}
 \eq
we come to the factorized form of the Lax matrix:
  \beq\label{q204}
  \begin{array}{c}
  \displaystyle{
L^\hbar(S(p,q),z)=\frac{\vth'(0)}{\vth(\hbar)}\,g(z+N\hbar,q)\,
 e^{P/c}\, g^{-1}(z,q)\,.
 }
 \end{array}
 \eq
The inverse gauge transformation provides (\ref{q365}):
  \beq\label{q205}
  \begin{array}{c}
  \displaystyle{
 L^{\rm RS}(z)=g^{-1}(z,q)L^\hbar(S(p,q),z)g(z,q)=\frac{\vth'(0)}{\vth(\hbar)}\,g^{-1}(z,q)g(z+N\hbar,q)\,
 e^{P/c}\,.
 }
 \end{array}
 \eq

 \subsection{Ruijsenaars-Schneider $M$-matrix in terms of $g(z)$}\label{Sect35}
 Let us compute the Ruijsenaars-Schneider
$M$-matrix using representation (\ref{q201}). Consider expansion of
the identity (\ref{q117}) near $\hbar$. Using (\ref{q101}) and
(\ref{q361}) we get in the $\hbar^{-1}$ order:
  \beq\label{q206}
  \begin{array}{c}
  \displaystyle{
 {\breve g}_2(0)=g_1(z){\mathcal O}_{12}\,g_1^{-1}(z)\,{\breve
 g}_2(0)\,.
 }
 \end{array}
 \eq
 It holds true by the following reason. Due to (\ref{q362}) ${\breve
 g}_{km}(0)=\psi_m$. Then the r.h.s. of (\ref{q206}) acquires the
 form:
  \beq\label{q207}
  \begin{array}{c}
  \displaystyle{
 g_1(z){\mathcal O}_{12}\,g_1^{-1}(z)\,{\breve
 g}_2(0)=\sum\limits_{i,j,k,l,m}g_{ik}(z)g_{kj}^{-1}(z)E_{ij}\otimes
 E_{lm}\, {\breve
 g}_{km}(0)=\sum\limits_{i,j,l,m}E_{ij}\delta_{ij}\otimes
 E_{lm}\psi_m\,.
 }
 \end{array}
 \eq
The latter is equal to ${\breve g}_2(0)$. For the $\hbar^0$ order of
the expansion of (\ref{q117}) we have:
  \beq\label{q208}
  \begin{array}{c}
  \displaystyle{
 \frac1N\,{\breve g}_2(0,q)\,r_{12}(z)=g_1'(z)\, \mathcal O_{12}\, {\breve g}_2(0)\, g_1^{-1}(z)+
 g_1(z)\, \mathcal O_{12}\,A_2\, g_1^{-1}(z)
 }
 \end{array}
 \eq
with matrix $A$ defined in (\ref{q361}). As in the previous
paragraph let us multiply both sides of (\ref{q208}) by
$g_2(N\hbar)e^{P_2/c}\vth'(0)/\vth(\hbar)$ from the left and compute
the trace over the second space. This provides
  \beq\label{q209}
  \begin{array}{c}
  \displaystyle{
 -M^\hbar(S(p,q),z)=\frac1N\,\tr_2(r_{12}(z)S_2(p,q))=g_1'(z)G_1g_1^{-1}(z)+g_1(z)F_1g_1^{-1}(z)\,,
 }
 \end{array}
 \eq
where
  \beq\label{q210}
  \begin{array}{c}
  \displaystyle{
 G_1=\tr_2\left({\mathcal O}_{12}\frac{\vth'(0)}{\vth(\hbar)}\,{\breve
 g}_2(0)g_2(N\hbar)\,
 e^{P_2/c}\right)
 }
 \end{array}
 \eq
and
  \beq\label{q211}
  \begin{array}{c}
  \displaystyle{
 F_1=\tr_2\left({\mathcal O}_{12}\frac{\vth'(0)}{\vth(\hbar)}\,A_2\,g_2(N\hbar)\,
 e^{P_2/c}\right).
 }
 \end{array}
 \eq
From (\ref{q201}) and the inverse gauge transformation (\ref{q205})
for the $M$-matrix
  \beq\label{q212}
  \begin{array}{c}
  \displaystyle{
 {M^{\rm RS}}'(z)=g^{-1}(z,q)M^\hbar(S(p,q),z)g(z,q)+g^{-1}(z,q){\dot g}(z,q)
 }
 \end{array}
 \eq
we get
  \beq\label{q213}
  \begin{array}{c}
  \displaystyle{
 -{M^{\rm RS}}'(z)=g^{-1}(z)g'(z)G+F-g^{-1}(z){\dot g}(z)\,,
 }
 \end{array}
 \eq
where
  \beq\label{q214}
  \begin{array}{c}
  \displaystyle{
 {\dot g}(z)=g'(z)\left(-N\,{\rm diag}(\dot q)+1_{N\times N}\sum\limits_k{\dot
 q}_k\right)-g(z){\dot D}^0(D^0)^{-1}\,.
 }
 \end{array}
 \eq
with ${\rm diag}(\dot q)$ being the diagonal matrix of the
velocities (\ref{q024}) defined as in (\ref{q3741}).

\begin{predl}
 The matrix ${M^{\rm RS}}'(z)$ in (\ref{q213}) coincides with the
 Ruijsenaars-Schneider $M$-matrix (\ref{q028}) up to unimportant term
 proportional to identity matrix.
\end{predl}
{\em\underline{{Proof}}}: Consider expansion of the
Ruijsenaars-Schneider Lax matrix near $z=0$
  \beq\label{q215}
  \begin{array}{c}
  \displaystyle{
 L^{\rm RS}=\frac{1}{z}\,L^{\rm RS}_{-1}+L^{\rm RS}_{0}+O(z)
 }
 \end{array}
 \eq
in two ways. First, from the definition (\ref{q025}):
  \beq\label{q216}
  \begin{array}{c}
  \displaystyle{
 L^{\rm RS}_{-1}=\rho\otimes\rho^T {\rm diag}(\dot q)\,,
 }
 \end{array}
 \eq
  \beq\label{q217}
  \begin{array}{c}
  \displaystyle{
 \left(L^{\rm RS}_{0}\right)_{ij}\stackrel{(\ref{q920})}{=}E_1(q_{ij}+\hbar)\,{\dot q}_j\,,
 }
 \end{array}
 \eq
where $\rho$ is the vector-column (\ref{q360}). The second way to
get (\ref{q215}) is to use (\ref{q205}):
  \beq\label{q218}
  \begin{array}{c}
  \displaystyle{
 L^{\rm RS}_{-1}=\frac{\vth'(0)}{\vth(\hbar)}\,{\breve
 g}(0)\,g(N\hbar)\,
 e^{P/c}\,,
 }
 \end{array}
 \eq
  \beq\label{q219}
  \begin{array}{c}
  \displaystyle{
 L^{\rm RS}_{0}=\frac{\vth'(0)}{\vth(\hbar)}\,{\breve
 g}(0)\,g'(N\hbar)\,e^{P/c}
 +\frac{\vth'(0)}{\vth(\hbar)}\,A\,g(N\hbar)\,e^{P/c}\,.
 }
 \end{array}
 \eq
Then, from (\ref{q210}) and (\ref{q218})
  \beq\label{q220}
  \begin{array}{c}
  \displaystyle{
 G_1=\tr_2\left({\mathcal O}_{12}(L^{\rm RS}_{-1})_2\right)\stackrel{(\ref{q216})}{=}
 \tr_2\left({\mathcal O}_{12}(\rho\otimes\rho^T)_2\, {\rm diag}(\dot q)_2\right)
 \stackrel{(\ref{q203})}{=}1_{N\times N}\sum_k {\dot q}_k\,.
 }
 \end{array}
 \eq
Plugging this into (\ref{q213}) we obtain
  \beq\label{q221}
  \begin{array}{c}
  \displaystyle{
  -{M^{\rm RS}}'(z)=Ng^{-1}(z)g'(z)\,{\rm diag}(\dot q)+F+{\dot
  D}^0(D^0)^{-1}\,.
 }
 \end{array}
 \eq
Notice that the last two terms are diagonal, so that the
non-diagonal part of ${M^{\rm RS}}'(z)$ is defined by the first term
only. The coincidence of the non-diagonal parts of ${M^{\rm
RS}}'(z)$ and (\ref{q028}) is due to identity (\ref{q9501}), which
comes from the non-relativistic limit of (\ref{q205}).

To complete the proof let us compute the matrix $F$ (\ref{q211}).
For this purpose substitute the matrix $L^{\rm RS}_{0}$ (\ref{q217})
into (\ref{q219})
  \beq\label{q222}
  \begin{array}{c}
  \displaystyle{
 \frac{\vth'(0)}{\vth(\hbar)}\,A\,g(N\hbar)\,e^{P/c}=L^{\rm RS}_{0}-\frac{\vth'(0)}{\vth(\hbar)}\,{\breve
 g}(0)\,g'(N\hbar)\,e^{P/c}
 }
 \end{array}
 \eq
and compute the last term in the r.h.s. by differentiating
 the identity (\ref{q218}) with respect to $\hbar$:
  \beq\label{q223}
  \begin{array}{c}
  \displaystyle{
 \p_\hbar L^{\rm RS}_{-1}=-E_1(\hbar)L^{\rm RS}_{-1}+N\frac{\vth'(0)}{\vth(\hbar)}\,{\breve
 g}(0)\,g'(N\hbar)\,e^{P/c}\,.
 }
 \end{array}
 \eq
From (\ref{q222}) and (\ref{q223}) we find
  \beq\label{q224}
  \begin{array}{c}
  \displaystyle{
 \frac{\vth'(0)}{\vth(\hbar)}\,A\,g(N\hbar)\,e^{P/c}=L^{\rm
 RS}_{0}-\frac1N\,\p_\hbar L^{\rm RS}_{-1}-\frac1N\,E_1(\hbar) L^{\rm
 RS}_{-1}\,.
 }
 \end{array}
 \eq
Plugging here (\ref{q216})-(\ref{q217}) yields
  \beq\label{q225}
  \begin{array}{c}
  \displaystyle{
 \left(\frac{\vth'(0)}{\vth(\hbar)}\,A\,g(N\hbar)\,e^{P/c}\right)_{il}=
 (L^{\rm RS}_{0})_{il}+\frac1N\,{\dot q}_l\sum\limits_{k\neq
 l}E_1(q_{lk}-\hbar)\,.
 }
 \end{array}
 \eq
Then, for the matrix $F$ (\ref{q211}) using (\ref{q203}) we obtain
  \beq\label{q226}
  \begin{array}{c}
  \displaystyle{
 F_{ij}=
 \delta_{ij}\left(\sum\limits_k(L^{\rm RS}_{0})_{ik}+\frac1N\,\sum\limits_{k\neq l}{\dot
 q}_l\,E_1(q_{lk}-\hbar)\right)\,.
 }
 \end{array}
 \eq
We now turn back to (\ref{q221}) and compute the diagonal part of
its r.h.s. The input (to $ii$-th diagonal element) of the first term
($Ng^{-1}(z)g'(z)\,{\rm diag}(\dot q)$) is evaluated from
(\ref{q9501}):
  \beq\label{q227}
  \begin{array}{c}
  \displaystyle{
{\dot q}_i\,E_1(z)-{\dot q}_i\sum\limits_{k\neq i}E_1(q_{ik})\,.
 }
 \end{array}
 \eq
The input of the $F$ matrix term  comes from (\ref{q226}) and
(\ref{q217}):
  \beq\label{q228}
  \begin{array}{c}
  \displaystyle{
 {\dot q}_i\,E_1(\hbar)+\sum\limits_{k\neq i}{\dot q}_k
 E_1(q_{ik}+\hbar)+\frac1N\,\sum\limits_{k,l:k\neq l}{\dot
 q}_l\,E_1(q_{lk}-\hbar)\,.
 }
 \end{array}
 \eq
At last, the input of the ${\dot   D}^0(D^0)^{-1}$ term is equal to
  \beq\label{q229}
  \begin{array}{c}
  \displaystyle{
 \sum\limits_{k\neq i}({\dot q}_i-{\dot q}_k)
 E_1(q_{ik})\,.
 }
 \end{array}
 \eq
Summing up (\ref{q227})-(\ref{q229}) we reproduce $-M^{\rm RS}_{ii}$
(\ref{q028}) except the last term from (\ref{q228}) which is
independent of $i$. It is proportional to the identity matrix, and
it has no affects on the Lax equations. $\blacksquare$

The $M$-matrix for rational Ruijsenaars-Schneider system has the form:
  \beq
 \begin{array}{c}
\displaystyle{ M^{ \rm RS}_{ij} =
-\left(1-\delta_{ij}\right)\left(\frac{1}{q_{i}-q_{j}} +
\frac{1}{Nz}\right)\dot{q}_{j}- }
\\ \ \\
\displaystyle{ - \delta_{ij}\left(\dot{q}_{i}\left(\frac{1}{Nz} +
\frac{1}{h}\right) + \sum\limits_{k \neq
i}^{N}\dot{q}_{k}\left(\frac{1}{q_{i}-q_{k}+h} -
\frac{1}{q_{i}-q_{k}}\right) \right) }\,.
 \end{array}
  \eq
  The $M$-matrix without spectral parameter can be obtained by
sending $z \rightarrow \infty$.

  \begin{example}
The $M$-matrix with spectral parameter for rational Ruijsenaars-Schneider
system, up to some unimportant terms proportional to identity
matrix, can be written in the following form:
  \beq M^{\rm RS}(z) = -
g^{-1}(z) g^{'}(z) {\rm{diag}}    (\dot{q}) - F -
\dot{D^{0}}(D^{0})^{-1},  \eq where  \beq
 \begin{array}{c}
\displaystyle{ g = \Xi (D^{0})^{-1} }\,,
\\ \ \\
\displaystyle{ F_{ij} = \delta_{ij}\sum\limits_{k=1}^{N}
\frac{\dot{q}_{k}}{q_{i}-q_{k}+h} }\,.
 \end{array}
  \eq
  The matrices $\Xi$, $D^{0}$ were defined in (\ref{q433}).
  \end{example}

  \begin{example}
The $M$-matrix without spectral parameter for rational Ruijsenaars-Schneider
system, up to some unimportant terms proportional to identity
matrix, can be written in the following form:
  \beq M^{\rm RS} = -
g^{-1}(z) g^{'}(z) {\rm{diag}}    (\dot{q}) - F -
\dot{D^{0}}(D^{0})^{-1}\,,
  \eq
  where
   \beq
 \begin{array}{c}
\displaystyle{ g = V (D^{0})^{-1} }\,,
\\ \ \\
\displaystyle{ F_{ij} = \delta_{ij}\sum\limits_{k=1}^{N}
\frac{\dot{q}_{k}}{q_{i}-q_{k}+h} }\,.
 \end{array}
  \eq
  The matrix $V$ was defined in (\ref{q4371}) and $D^{0}$ in
(\ref{q433}).
 \end{example}
\vskip5mm
The $M$-matrix for trigonometric Ruijsenaars-Schneider system with spectral parameter has the following form:
 \beq
 \begin{array}{c}
\displaystyle{
M_{ij}^{\rm RS}(z) = -(1-\delta_{ij})(\coth(q_{i}-q_{j}) + \coth(Nz))\dot{q}_{j} -
}
\\ \ \\
\displaystyle{
-\delta_{ij}
 \left(\dot{q}_{i}(\coth(Nz) + \coth(\hbar)) +
 \sum\limits_{k \neq i}^{N} \dot{q}_{k}(\coth(q_{i}-q_{k}+\hbar) -
 \coth(q_{i}-q_{k}))\right)\,.
}
 \end{array}
 \eq
The corresponding $M$-matrix without spectral parameter:
 \beq
 \begin{array}{c}
\displaystyle{
M_{ij}^{\rm RS} = -(1-\delta_{ij})\frac{\dot{q}_{j}}{\sinh(q_{i}-q_{j})} -
}
\\ \ \\
\displaystyle{
 -\delta_{ij} \left(\dot{q}_{i} \coth(\hbar) +
\sum\limits_{k \neq i}^{N}\left(\dot{q}_{k}(\coth(q_{i}-q_{k}+\hbar)
- \coth(q_{i}-q_{k}) \right)\right)\,.
 }
 \end{array}
 \eq

 \begin{example}
The $M$-matrix with spectral parameter for the trigonometric
Ruijsenaars-Schnei\-der system, up to some unimportant terms
proportional to identity matrix, can be written in form:
  \beq
M^{\rm RS}(z) = - g^{-1}(z)g^{'}(z) {\rm diag}(\dot{q}) - F -
\dot{D}^{0}(D^{0})^{-1},
 \eq
  where
  \beq
 \begin{array}{c}
 \displaystyle{
 g = \tilde{\Xi} (D^{0})^{-1}\,,
 }
\\ \ \\
\displaystyle{ F_{ij} =
\delta_{ij}\sum\limits_{k=1}^{N}\coth(q_{i}-q_{k}+\hbar)\dot{q}_{k}\,.
}
 \end{array}
 \eq The matrices $\tilde{\Xi}$ and $D^{0}$ were defined in
(\ref{q423}).
 \end{example}

 \begin{example}

The $M$-matrix without spectral parameter for the trigonometric
Ruijsenaars-Schneider system, up to some unimportant terms
proportional to identity matrix, can be written in form:  \beq
M^{\rm RS} = - g^{-1}(z)g^{'}(z) {\rm diag}(\dot{q}) - F -
\dot{D}^{0}(D^{0})^{-1},  \eq where  \beq
 \begin{array}{c}
\displaystyle{ g = \tilde{V} (\tilde{D}^{0})^{-1}\,, }
\\ \ \\
\displaystyle{ F_{ij} =
\delta_{ij}\sum\limits_{k=1}^{N}\coth(q_{i}-q_{k}+\hbar)\,\dot{q}_{k}\,.
}
 \end{array}
 \eq The matrices $\tilde{V}$ and $\tilde{D}^{0}$ were defined in
(\ref{q4261}) and (\ref{q4262}).

 \end{example}


\section{Schlesinger transformation}\label{sect4}
\setcounter{equation}{0}
 %
%
%
%
%
%
%

In this Section we will show that the Lax pair of the Calogero-Moser
model (\ref{q029})-(\ref{q030}) is naturally obtained from the
Schlesinger transformation  generated by the intertwining matrix
(\ref{q52})-(\ref{q54}).

The Schlesinger transformation
\cite{Schl1,Schl2,Schl3,Arinkin1,Arinkin2} is a (singular in the
local coordinate $z$) gauge transformation
  \beq\label{q701}
  \begin{array}{c}
  \displaystyle{
 A(z)\rightarrow hA(z)h^{-1}-\p_zhh^{-1}
 }
 \end{array}
 \eq
of (the $z$ component of) a connection, which changes its residues.
For example, in the simplest case the scalar connection on ${\mathbb
CP}^1$ $A(z)=\p_z+\nu_0/z$, where $\nu_0$ is a constant, is
transformed via (\ref{q701}) with $h=z$ as
$\nu_0\rightarrow\nu_0-1$. Similarly, on the elliptic curve the
scalar connection $A(z)=\p_z+\nu_0E_1(z)$, where $E_1(z)$ is
(\ref{q912}) is transformed via (\ref{q701}) with $h=\vth(z)$ as
$\nu_0\rightarrow\nu_0-1$ as well. As we know from (\ref{q9501}) the
non-trivial part (corresponding to the non-zero coupling constant
$\nu$) of the Lax matrix (\ref{q0297}) has form of a pure gauge
connection along the coordinate $z$ on the elliptic curve. We are
going to treat it as a result of the Schlesinger like
transformation. To make sense of a connection along  the spectral
parameter $z$ we should proceed to the monodromy preserving
equations.

\noindent{\bf Classical Painlev\'e-Calogero correspondence.} As is
known from \cite{LO} the Lax pair (\ref{q029})-(\ref{q030})
satisfies not only the Lax equation ${\dot L}=[L,M]$ but also the
zero curvature condition
  \beq\label{q702}
  \begin{array}{c}
  \displaystyle{
 2\pi\imath\frac{d}{d\tau}L-\frac{d}{dz}M=[L,M]\,.
 }
 \end{array}
 \eq
More precisely, the $M$-matrix (\ref{q030}) should be shifted by the
identity matrix multiplied by $\p_\tau\log\vth(z)$: $M\rightarrow
M+1_N\p_\tau\log\vth(z)$ in order to compensate $2\pi\imath\p_\tau
E_1(z)$ coming from the first term in the l.h.s. of (\ref{q702}).
Then (\ref{q702}) is equivalent (identically in $z$) to the higher
Painlev\'e equations
  \beq\label{q703}
  \begin{array}{c}
  \displaystyle{
 (2\pi\imath)^2\frac{d^2{q}_i}{d\tau^2}=\nu^2\sum\limits_{k\neq i}\wp'(q_{ik})\,.
 }
 \end{array}
 \eq
This system of equations is treated as non-autonomous version of the
Calogero-Moser equations of motion (\ref{q032}) in the sense that
the elliptic moduli $\tau$ (entering the r.h.s. of (\ref{q703})
explicitly) plays the role of the time variable. Technically,
equivalence of (\ref{q702}) and (\ref{q703}) is similar to
derivation of the Lax equations for the Calogero-Moser model
together with the usage of $2\pi\imath\p_\tau L=\frac{d}{dz}M$. The
latter follows from the heat equations
(\ref{q9292})-(\ref{q9293})\footnote{Let us remark that the property
$2\pi\imath\p_\tau L=\frac{d}{dz}M$ is gauge dependent, so that the
gauge choice including $D^0$ matrix is important here.}.

Another important argument is that all models connected by the
symplectic Hecke correspondence satisfy the property of the
Painlev\'e-Calogero correspondence as well \cite{LOZ13}. So that the
gauge transformed Lax pair again satisfies not only the Lax equation
but also the zero-curvature condition (\ref{q702}) if the gauge
transformation is given by the modification of the underlying
bundle.

Then we may perform the following procedure. Consider the Lax matrix
of the Calogero-Moser model with the coupling constant $\nu_0$:
  \beq\label{q704}
  \begin{array}{c}
  \displaystyle{
 L_0=P+N\nu_0g^{-1}g'\,.
 }
 \end{array}
 \eq
 At
first, perform the gauge transformation generated by $g$-matrix.
Secondly, transform the Lax matrix into the connection by adding
$\p_z$. Thirdly, perform the inverse gauge transformation generated
by $g^{-1}$-matrix. At last, reduce the connection to the Lax
matrix. The validity of the second and the last steps is guaranteed
by the Painlev\'e-Calogero correspondence. Schematically, the
procedure is as follows:
  \beq\label{q705}
  \begin{array}{c}
  \displaystyle{
 L_0\rightarrow gL_0g^{-1}\rightarrow \p_z+gL_0g^{-1}\rightarrow \p_z+L_0+g^{-1}g'\rightarrow L_0+g^{-1}g'=P+(N\nu_0+1)g^{-1}g'\,.
 }
 \end{array}
 \eq
As a result we get the same Lax matrix with the coupling constant
shifted as $\nu_0\rightarrow\nu_0+1/N$.

\noindent{\bf Calogero-Moser $M$-matrix in terms of $g(z)$.} The
described above procedure is a way to get the non-trivial part of
the Calogero-Moser Lax matrix in the form of the pure gauge
connection. Let us repeat all the steps to get the $M$-matrix. For
convenience let us set $\nu_0=0$. Then the initial $M$-matrix equals
zero since it corresponds to the free model. The analogue of
(\ref{q705}) is as follows:
  \beq\label{q706}
  \begin{array}{c}
  \displaystyle{
 M_0=0\rightarrow -{\dot g}g^{-1}\rightarrow 2\pi\imath\p_\tau-{\dot g}g^{-1}
 \rightarrow 2\pi\imath\p_\tau-g^{-1}{\dot g}-g^{-1}\frac{d}{d\tau}\,g\rightarrow
 M
 }
 \end{array}
 \eq
where
  \beq\label{q711}
  \begin{array}{c}
  \displaystyle{
 M=g^{-1}\frac{d}{d\tau}\,g-g^{-1}\frac{d}{dt}\,g\,.
 }
 \end{array}
 \eq
Both derivatives are the full derivatives, i.e. they include
differentiation with respect to explicit and implicit dependencies
on these variables. The implicit one is contained in $q_i(t)$ or
$q_i(\tau)$. The relation between momenta and velocities comes from
the Hamiltonian equations with the Hamiltonian function being
computed from $(1/2)\tr L^2$. Notice that at the first and the
second stages of (\ref{q705}) we have $p_i={\dot q}_i$, while on the
last two stages an additional terms appear coming from the diagonal
part of the $g^{-1}g'$:
  \beq\label{q7101}
  \begin{array}{c}
  \displaystyle{
 {\rm diag}(q)_t=P\,,\quad {\rm diag}(q)_\tau=P-\frac{1}{N}\,d\,.
 }
 \end{array}
 \eq
where
  \beq\label{q7121}
  \begin{array}{c}
  \displaystyle{
d_i=\sum\limits_{k\neq i}E_1(q_{ik})\,.
 }
 \end{array}
 \eq
So that
  \beq\label{q710}
  \begin{array}{c}
  \displaystyle{
 {\rm diag}(q)_\tau-{\rm diag}(q)_t=-\frac1N\,d\,.
 }
 \end{array}
 \eq
The latter relation explains how to compute $M$-matrix via
(\ref{q711}).

Introduce notations:
  \beq\label{q712}
  \begin{array}{c}
  \displaystyle{
Ng^{-1}g'=l(z)\,,
 }
 \end{array}
 \eq
  \beq\label{q7122}
  \begin{array}{c}
  \displaystyle{
l_{ii}(z)=E_1(z)-\sum\limits_{k\neq i}E_1(q_{ik})=E_1(z)-d_i\,.
 }
 \end{array}
 \eq
From (\ref{q712}) we also have
  \beq\label{q713}
  \begin{array}{c}
  \displaystyle{
 Ng^{-1}g''=\p_zl(z)+\frac1N\,l^2(z)\,.
 }
 \end{array}
 \eq

\begin{predl}
 The matrix $M(z)$ (\ref{q711}) with the relation (\ref{q710}) coincides with the
 Calogero-Moser $M$-matrix (\ref{q030}) up to unimportant terms
 proportional to the identity matrix.
\end{predl}
{\em\underline{{Proof}}}:

From the explicit form of $g$ (\ref{q52}) and (\ref{q710}) we get
  \beq\label{q714}
  \begin{array}{c}
  \displaystyle{
 M=\frac{N}{2}\,g^{-1}g''-2\pi\imath D^{-1}\p_\tau
 D-D^{-1}{\dot D}\left.\right|_{{\dot q}_i=-d_i/N}
 -Ng^{-1}g'({\rm diag}(q)_\tau-{\rm diag}(q)_t)=
 }
 \\ \ \\
  \displaystyle{
=\frac{N}{2}\,g^{-1}g''-2\pi\imath D^{-1}\p_\tau
 D+\frac1N\,D^{-1}{\dot D}\left.\right|_{{\dot q}_i=d_i}
 +g^{-1}g'd\,.
 }
 \end{array}
 \eq

\underline{Non-diagonal part:}

  \beq\label{q715}
  \begin{array}{c}
  \displaystyle{
 (Ng^{-1}g'')_{ij}=\p_zl_{ij}+\frac1N\,l_{ij}(l_{ii}+l_{jj})+\frac1N\sum\limits_{k\neq
 i,j}l_{ik}l_{kj}\,.
 }
 \end{array}
 \eq
Using (\ref{q715}) and
  \beq\label{q716}
  \begin{array}{c}
  \displaystyle{
 \p_zl_{ij}\stackrel{(\ref{q917})}{=}\phi(z,q_{ij})(E_1(z+q_{ij})-E_1(z))
 }
 \end{array}
 \eq
together with
  \beq\label{q717}
  \begin{array}{c}
  \displaystyle{
 l_{ik}l_{kj}\stackrel{(\ref{q915})}{=}\phi(z,q_{ij})(E_1(z)+E_1(q_{ik})+E_1(q_{kj})-E_1(z+q_{ij}))
 }
 \end{array}
 \eq
we get
  \beq\label{q718}
  \begin{array}{c}
  \displaystyle{
 (\frac{N}{2}\,g^{-1}g'')_{ij}=\frac1N\,f(z,q_{ij})-\frac1N\,l_{ij}d_j\,,
 }
 \end{array}
 \eq
which means that for $i\neq j$ the statement of the Proposition
indeed holds true.

\underline{Diagonal part:}

The inputs coming from (\ref{q714}) are as follows. From
(\ref{q713}) using (\ref{q916}) we find
  \beq\label{q719}
  \begin{array}{c}
  \displaystyle{
 \frac12(Ng^{-1}g'')_{ii}=\frac{1}{2N}\left( E_1^2(z)-E_2(z)
 \right)+\frac{1}{2N}\,d_i^2-\frac{1}{2N}\sum\limits_{k\neq
 i}E_2(q_{ik})-\frac{1}{N}\,E_1(z)d_i\,.
 }
 \end{array}
 \eq
Next,
  \beq\label{q720}
  \begin{array}{c}
  \displaystyle{
 -2\pi\imath D^{-1}_i\p_\tau D_i=-\frac12\sum\limits_{k\neq i}\p_\tau\log\vth(q_{ik})
 =-\frac12\sum\limits_{k\neq i}E_1^2(q_{ik})+\frac12\sum\limits_{k\neq
 i}E_2(q_{ik})\,.
 }
 \end{array}
 \eq
Next,
  \beq\label{q721}
  \begin{array}{c}
  \displaystyle{
 \frac1N\,D_i^{-1}{\dot D}_i\left.\right|_{{\dot q}_i=d_i}
 =\frac1N\sum\limits_{k\neq i} (d_i-d_k)E_1(q_{ik})\,.
 }
 \end{array}
 \eq
Finally,
  \beq\label{q722}
  \begin{array}{c}
  \displaystyle{
 \left(g^{-1}g'd\right)_{ii}=\frac1N\,E_1(z)d_i-\frac1N\,d_i^2\,.
 }
 \end{array}
 \eq
Summarizing (\ref{q719})-(\ref{q722}) for the diagonal part of
(\ref{q714}) we get
  \beq\label{q723}
  \begin{array}{c}
  \displaystyle{
 M_{ii}=\frac1N\p_\tau\log\vth(z)+\frac{N-1}{2N}\sum\limits_{k\neq i}E_2(q_{ik})
 +\frac{1}{2N}\,d_i^2-\frac12\sum\limits_{k\neq i}E_1^2(q_{ik})-\frac1N\sum\limits_{k\neq i}
 d_kE_1(q_{ik})\,.
 }
 \end{array}
 \eq
Introduce notation
  \beq\label{q724}
  \begin{array}{c}
  \displaystyle{
 {\sum\limits_{kl}}''=\sum\limits_{k,l:k\neq i,l\neq i, k\neq l}\,.
 }
 \end{array}
 \eq
Since
  \beq\label{q725}
  \begin{array}{c}
  \displaystyle{
 \sum\limits_{k\neq i}d_kE_1(q_{ik})=-\sum\limits_{k\neq i}E_1^2(q_{ik})+{\sum\limits_{kl}}''E_1(q_{ik})E_1(q_{kl})
 }
 \end{array}
 \eq
and
  \beq\label{q726}
  \begin{array}{c}
  \displaystyle{
 d_i^2=\sum\limits_{k\neq i}E_1^2(q_{ik})+{\sum\limits_{kl}}''E_1(q_{ik})E_1(q_{il})
 }
 \end{array}
 \eq
the expression (\ref{q723}) acquires the form:
  \beq\label{q727}
  \begin{array}{c}
  \displaystyle{
 M_{ii}=\frac1N\p_\tau\log\vth(z)+\frac{N-1}{2N}\sum\limits_{k\neq
 i}E_2(q_{ik})-
 }
 \\ \ \\
  \displaystyle{
 -\frac{N-3}{2N}\sum\limits_{k\neq i}E_1^2(q_{ik})+\frac{1}{2N}{\sum\limits_{kl}}''E_1(q_{ik})E_1(q_{il})
 -\frac1N{\sum\limits_{kl}}''E_1(q_{ik})E_1(q_{kl})\,.
 }
 \end{array}
 \eq
Consider the following sums
  \beq\label{q728}
  \begin{array}{c}
  \displaystyle{
 \Delta_i={\sum\limits_{kl}}''(E_1(q_{ik})+E_1(q_{kl})+E_1(q_{li}))^2\,.
 }
 \end{array}
 \eq
Due to
  \beq\label{q729}
  \begin{array}{c}
  \displaystyle{
 {\sum\limits_{kl}}''E_1^2(q_{kl})=\sum\limits_{k,l:k\neq
 l}E_1^2(q_{kl})-2\sum\limits_{k\neq i}E_1^2(q_{ik})
 }
 \end{array}
 \eq
and
  \beq\label{q730}
  \begin{array}{c}
  \displaystyle{
 {\sum\limits_{kl}}''E_1^2(q_{ik})=(N-2)\sum\limits_{k\neq i}E_1^2(q_{ik})
 }
 \end{array}
 \eq
we have
  \beq\label{q731}
  \begin{array}{c}
  \displaystyle{
 \Delta_i=
  }
 \\ \ \\
  \displaystyle{
 =\sum\limits_{k,l:k\neq l}E_1^2(q_{kl})+2(N-3)\sum\limits_{k\neq i}E_1^2(q_{ik})
 +4{\sum\limits_{kl}}''E_1(q_{ik})E_1(q_{kl})-2{\sum\limits_{kl}}''E_1(q_{ik})E_1(q_{il})\,.
 }
 \end{array}
 \eq
Then expression (\ref{q727}) is simplified as follows:
  \beq\label{q732}
  \begin{array}{c}
  \displaystyle{
 M_{ii}=\frac1N\p_\tau\log\vth(z)+\frac{1}{4N}\sum\limits_{k,l:k\neq l}E_1^2(q_{kl})+\frac{N-1}{2N}\sum\limits_{k\neq
 i}E_2(q_{ik})-\frac{1}{4N}\,\Delta_i\,.

 }
 \end{array}
 \eq
Notice that the first and the second terms are independent of index
$i$. They provide the term proportional to the identity matrix. The
sum $\Delta_i$ (\ref{q728}) can be written in a different way using
(\ref{q9541}). Plugging for each term of the sum (\ref{q728}) the
r.h.s. of (\ref{q9541}) we obtain:
  \beq\label{q733}
  \begin{array}{c}
  \displaystyle{
 \Delta_i=(N-1)(N-2)\frac{\vth'''(0)}{\vth'(0)}+2(N-3)\sum\limits_{k\neq i}E_2(q_{ik})
 +\sum\limits_{k,l:k\neq l}E_2(q_{kl})
 }
 \end{array}
 \eq
Then for the diagonal part of the $M$-matrix (\ref{q732}) we get
  \beq\label{q734}
  \begin{array}{c}
  \displaystyle{
 M_{ii}=\frac1N\p_\tau\log\vth(z)-\frac{(N-1)(N-2)}{4N}\frac{\vth'''(0)}{\vth'(0)}+
 \frac{1}{4N}\sum\limits_{k,l:k\neq l}(E_1^2(q_{kl})-E_2(q_{ik}))+
 }
 \\ \ \\
  \displaystyle{
 +\frac{1}{N}\sum\limits_{k\neq i}E_2(q_{ik})\,.
 }
 \end{array}
 \eq
All terms in the upper line of (\ref{q734}) are independent of index
$i$, and the lower line is the diagonal part of (\ref{q030}) with
$\nu=1/N$. $\blacksquare$

\noindent{\bf Examples.}

 \begin{example}
The $M$-matrix of the rational Calogero-Moser model
 \beq
M_{ij} =  \delta_{ij}\left(\sum\limits_{k \neq
i}^{N}\frac{\nu}{(q_{i}-q_{k})^{2}}\right) -
(1-\delta_{ij})\frac{\nu}{(q_{i}-q_{j})^{2}}
 \eq
up to sum unimportant terms proportional to the identity matrix can be written as follows:
 \beq
M = \nu\left(\frac{1}{2}g^{-1}g^{''} + g^{-1}g^{'}d +
(D^{0})^{-1}\dot{D^{0}}|_{\dot{q}_{i} = d_{i}}\right)\,,
 \eq
where
 \beq
 \begin{array}{c}
    \displaystyle{
    g = \Xi (D^{0})^{-1}\,,
    }
    \\ \ \\
    \displaystyle{
    d_{ij} = \delta_{ij} d_{i} = \delta_{ij} \sum\limits_{k \neq i}^{N}\frac{1}{q_{i} -
    q_{k}}\,.
    }
 \end{array}
 \eq
The matrices $\Xi$, $D^{0}$ were defined in (\ref{q433}).
 \end{example}

 \begin{example}
The $M$-matrix of the rational Calogero-Moser model
 \beq
M_{ij} =  \delta_{ij}\left(\sum\limits_{k \neq
i}^{N}\frac{\nu}{(q_{i}-q_{k})^{2}}\right) -
(1-\delta_{ij})\frac{\nu}{(q_{i}-q_{j})^{2}}
 \eq
up to sum unimportant terms proportional to the identity matrix can be written as follows:
 \beq
M = \nu\left(\frac{1}{2}g^{-1}g^{''} + g^{-1}g^{'}d +
(D^{0})^{-1}\dot{D^{0}}|_{\dot{q}_{i} = d_{i}}\right),
 \eq
where
 \beq
 \begin{array}{c}
    \displaystyle{
    g = V (D^{0})^{-1}\,,
    }
    \\ \ \\
    \displaystyle{
    d_{ij} = \delta_{ij} d_{i} = \delta_{ij} \sum\limits_{k \neq i}^{N}\frac{1}{q_{i} -
    q_{k}}\,.
    }
 \end{array}
 \eq
The matrix $V$ was defined in (\ref{q4371}) and $D^{0}$ in
(\ref{q433}).
 \end{example}

 \begin{example}
The $M$-matrix of the trigonometric Calogero-Moser model
 \beq
M_{ij} = \delta_{ij} \left(\sum\limits_{k \neq i}^{N}\frac{\nu}
{\sinh^{2}(q_{i}-q_{k})}\right) -
\nu(1-\delta_{ij})\frac{1}{\sinh^{2}(q_{i}-q_{j})}
 \eq
up to some unimportant terms proportional to the identity matrix can
be written in form:
 \beq
M =\nu\left( \frac{1}{2} g^{-1} g^{''} + g^{-1}g^{'} d
+(\tilde{D}^{0})^{-1}\dot{\tilde{D}^{0}}|_{\dot{q}_{i} =
d_{i}}\right)\,,
 \eq
where
 \beq
 \begin{array}{c}
\displaystyle{ g = \tilde{\Xi} (\tilde{D}^{0})^{-1} }\,,
\\ \ \\
\displaystyle{ d_{ij} = \delta_{ij}d_{i} = \delta_{ij}
\left(\sum\limits_{k \neq i}^{N}\coth(q_{i}-q_{k})\right) -
\left(N-2\right)\,.
 }
 \end{array}
 \eq
The matrices $\tilde{\Xi}$ and $\tilde{D}^{0}$ were defined in
(\ref{q423}).
 \end{example}

 \begin{example}
The $M$-matrix of the trigonometric Calogero-Moser model
 \beq
M_{ij} = \delta_{ij} \left(\sum\limits_{k \neq i}^{N}\frac{\nu}
{\sinh^{2}(q_{i}-q_{k})}\right) -
\nu(1-\delta_{ij})\frac{\coth(q_{i}-q_{j})}{\sinh(q_{i}-q_{j})}
 \eq
up to some unimportant terms proportional to the identity matrix can
be written in form:
 \beq
M = \nu\left(\frac{1}{2} g^{-1} g^{''} + g^{-1}g^{'} d
+(\tilde{D}^{0})^{-1}\dot{\tilde{D}^{0}}|_{\dot{q}_{i} =
d_{i}}\right)\,,
 \eq
where
 \beq
 \begin{array}{c}
\displaystyle{ g = \tilde{V} (\tilde{D}^{0})^{-1} }\,,
\\ \ \\
\displaystyle{ d_{ij} = \delta_{ij}d_{i} = \delta_{ij}
\sum\limits_{k \neq i}^{N}\coth(q_{i}-q_{k})\,. }
 \end{array}
 \eq
The matrices $\tilde{V}$ and $\tilde{D}^{0}$ were defined in
(\ref{q4261})-(\ref{q4262}).
 \end{example}


\section{Classical root systems}
\setcounter{equation}{0}
 In this Section we propose factorization
formulae for the rational Calogero-Moser systems associated with
classical root systems $D_{N}$, $C_{N}$, $B_{N}$. As was mentioned
in the Introduction, in case when there is no spectral parameter the
factorization of the $A_N$ Lax matrix takes the form (\ref{q018}).
It is due to the fact that
 \beq\label{q5011}
\begin{array}{c}
    \displaystyle{
        V'(z)=C_0V(z)\,,
    }
\end{array}
\eq
where $V(z)$ is the Vandermonde matrix (\ref{q4371}) and
 \beq\label{q507}
 \begin{array}{c}
 (C_{0})_{ij}= \left\{\begin{array}{l}
 i-1,  i=j+1,\\
 0,\  \hbox{otherwise}\,.
  \end{array}\right.
 \end{array}
\eq
Below we suggest analogues of (\ref{q018}) for the models related to
$D_{N}$, $C_{N}$, $B_{N}$ root systems. The proofs are given in the
Appendix B.
\subsection{Calogero-Moser model associated with classical root systems}
The rational $BC_N$ model with two coupling constants is described
by the following Hamiltonian:
 \beq
\label{q501}
\begin{array}{c}
 \displaystyle{
H^{\rm CM} =\frac{1}{2} \sum_{i=1}^{N} p_{i}^{2} -
\left(\sum\limits_{j \leq i}^{N}
(\frac{m_{2}^{2}}{(q_{i}-q_{j})^{2}}+\frac{m_{2}^{2}}{(q_{i}+q_{j})^{2}})
+\sum\limits_{i=1}^{N} \frac{m^{2}}{(2q_{i})^{2}}\right)= }
\\ \ \\
 \displaystyle{
=\frac{1}{2} \sum_{i=1}^{N} p_{i}^{2} - \left(\sum\limits_{j \leq
i}^{N}
(\frac{m_{2}^{2}}{(q_{i}-q_{j})^{2}}+\frac{m_{2}^{2}}{(q_{i}+q_{j})^{2}})
+\sum\limits_{i=1}^{N} \frac{m_{4}^{2}}{(2q_{i})^{2}} +
 \sum\limits_{i=1}^{N} \frac{m_{1}^{2}}{q_{i}^{2}}\right)\,,
}
 \end{array}
 \eq
where\footnote{There are two independent constants in (\ref{q501}).}
$m^{2} = m_{4}^{2}+4m_{1}^{2}$. Its Lax matrix is of
$(2N+1)\times(2N+1)$ size \cite{OP1,OP2,DP}:
 \beq\label{q502}
\begin{array}{c}
    \displaystyle{
        L^{\rm CM}(m_{1},m_{2},m_{4})=\left(\begin{array}{ccc}
            A & B & C\\
            -B&-A&-C\\
            -C^{T}&C^{T}& 0
        \end{array}\right)
    }
\end{array}
 \eq
where the blocks $A$, $B$ are $N\times N$ matrices and $C$ is
$N$-dimensional column-vector:
 \beq\label{q503}
\begin{array}{c}
    \displaystyle{
        A_{ij}=\delta_{ij}\left(p_{i} - \frac{\sqrt{2}m_{4}}{2q_{i}} - \frac{\sqrt{2}m_{1}}{q_{i}} -
        m_{2}\sum\limits_{k \neq i}^{N}(\frac{1}{q_{i}-q_{k}} +
        \frac{1}{q_{i}+q_{k}})\right)+(1-\delta_{ij})\frac{m_{2}}{q_{i}-q_{j}}\,,
    }
    \\ \ \\
    \displaystyle{
        B_{ij}=(1-\delta_{ij})\frac{m_{2}}{q_{i}+q_{j}} +\delta_{ij}\frac{\sqrt{2}m_{4}}{2q_{i}}\,,
    }
    \\ \ \\
    \displaystyle{
        C_{i} = \frac{m_{1}}{q_{i}}\,.
    }
\end{array}
 \eq
The Lax matrix (\ref{q502}) obeys the Lax equations if $m_1(m_1^2
-2m_2^2+\sqrt{2}m_2 m_4) = 0$. It reduces to the classical root
systems $D_{N}$, $C_{N}$ and $B_{N}$ by following choice of the
coupling constants:
 \beq\label{q504}
\begin{array}{l}
    \displaystyle{
        D_{N}:\:\:\:\:\: m = 0\ \ (m_1 = 0,\:\: m_4 = 0)\,,
    }
    \\ \ \\
    \displaystyle{
        C_{N}:\:\:\:\:\:\: m^{2} = m_{4}^{2}\ \ (m_1 = 0)\,, \:\:\:\:\:\:\:\:\:\:\:\:\:\:\:\:
    }
    \\ \ \\
    \displaystyle{
        {B_{N}:} \:\:\:\:\:\:\:m^{2} = 4m_{1}^{2}\ \ (m_4 = 0\,,\:\: m_1^2 = 2m_2^2)\,.
    }
\end{array}
 \eq
  Notice that for $C_{N}$ and $D_{N}$ cases the Lax matrix is
effectively of dimension $2N$, therefore we will consider such
matrices as Mat($2N$,$\mathbb{C}$)-valued. In fact, more general Lax
pairs are know for the $BC_N$ Calogero-Moser and
Ruijsenaars-Schneider models
\cite{Feher2,Feher11,Feher12,Feher13,Feher14}, which are of size
$2N\times 2N$, and have no restrictions on the coupling constants.
Moreover, the group-theoretical construction underlying these Lax
pairs implies a kind of factorization formulae as well. Possible
applications of those results to quantum-classical dualities will be
discussed
 in our future publications.
\subsection{Factorization formulae for classical root systems}
\subsubsection*{Factorization for $C_{N}$ and $D_{N}$ root systems}
Introduce the following notations for $2N\times 2N$ matrices:
 \beq\label{q505}
\begin{array}{c}
    \displaystyle{
        D^{0}_{ij} = \delta_{ij} \left\{\begin{array}{l}
            2q_{i}\prod\limits_{k \neq i}^{N}((q_{i}-q_{k})(q_{i}+q_{k}))\,,\  i \leq N\,,\\
            -2q_{i-N}\prod\limits_{k \neq i-N}^{N}((q_{i-N}-q_{k})(q_{i-N}+q_{k}))\,,\  N+1 \leq i \leq
            2N\,,
        \end{array}\right.
    }
\end{array}
 \eq
 \begin{equation}
\label{q506} V_{ij} =  \left\{ \begin{array}{l}
q_{j}^{i-1},\  j \leq N\,,\\
(-q_{j-N})^{i-1},\  N+1 \leq j \leq 2N
 \end{array}\right.
 \end{equation}
and
 \begin{equation}
\label{q508} \tilde{C}_{ij} = \left\{ \begin{array}{l}
1,\  i=j+1\,,\ i-\hbox{even}\\
0,\  \hbox{otherwise}
 \end{array}\right.
 \end{equation}
The Lax matrices (\ref{q502}) for the $C_{N}$ and $D_{N}$ cases
(\ref{q504})  admit the following factorization formula:
 \begin{equation}
\label{q509} L^{\rm CM}(m_{2},m_{4},0) = P -
D^{0}V^{-1}(m_{2}C_{0}-(m_{2}-\sqrt{2}m_{4})\tilde{C})V(D^{0})^{-1},
 \end{equation}
where $C_0$ is the one (\ref{q507}) but of the size $2N\times 2N$,
and
 \begin{equation}
\label{q510}
 P_{ij} =\delta_{ij} \left\{ \begin{array}{l}
p_{i}\,,\  i \leq N\,,\\
-p_{i-N}\,,\ N+1 \leq i \leq 2N\,.
 \end{array}\right.
 \end{equation}
For the choice $m_{4} =0$ (\ref{q509}) reproduces the Lax matrix for
$D_{N}$ root system, otherwise we get the $C_{N}$ case.

\subsubsection*{Factorization for $B_{N}$ root system}
Let us introduce the notations for $(2N+1)\times(2N+1)$ matrices:
 \beq\label{q511}
 \begin{array}{c}
    \displaystyle{
        D^{0}_{ij} = \delta_{ij} \left\{ \begin{array}{l}
            \sqrt{2}q_{i}^{2}\prod\limits_{k \neq i}^{N}((q_{i}-q_{k})(q_{i}+q_{k}))\,,\ \: i \leq N\,,\\
            \sqrt{2}q_{i-N}^{2}\prod\limits_{k \neq i-N}^{N}((q_{i-N}-q_{k})(q_{i-N}+q_{k}))\,,\ \: N+1 \leq i \leq 2N\,, \\
            \prod\limits_{k=1}^{N}(-q_{k}^{2})\,,\ \:i=2N+1\,,
         \end{array}\right.
    }
 \end{array}
 \eq
 \begin{equation}
\label{q512} V_{ij} =  \left\{ \begin{array}{l}
q_{j}^{i-1}\,,\  j \leq N\,,\\
(-q_{j-N})^{i-1}\,,\  N+1 \leq j \leq 2N\,,\\
\delta_{i,1}\,,\ j=2N+1
 \end{array}\right.
 \end{equation}
and
 \begin{equation}
\label{q514} \tilde{C}_{ij} = \left\{ \begin{array}{l}
1,\  i=j+1,\: i-\hbox{even}\\
0,\  \hbox{otherwise}
 \end{array}\right.; \;\;\;\;i,j=1,..,2N+1\,.
 \end{equation}
The Lax matrix (\ref{q502}) for the $B_N$ case (\ref{q504}) admits
the following factorization formula:
 \begin{equation}
\label{q515} L^{\rm CM}(m_{2},0,\sqrt{2}m_{2}) =
P-m_{2}D^{0}V^{-1}(C_{0}+\tilde{C})V(D^{0})^{-1},
 \end{equation}
where $C_0$ is the one (\ref{q507}) but of the size
$(2N+1)\times(2N+1)$, and
 \begin{equation}
\label{q516} P_{ij} =\delta_{ij} \left\{ \begin{array}{l}
p_{i}\,,\  i \leq N\,,\\
-p_{i-N}\,,\ N+1 \leq i \leq 2N\,,\\
0\,,\ i=2N+1\,.
 \end{array}\right.
 \end{equation}


\section{Appendix A}

\def\theequation{A.\arabic{equation}}
\setcounter{equation}{0}

\paragraph{Finite dimensional representation of Heisenberg group.}
Instead of the standard basis in ${\rm Mat}_{ N}$ the following one
is widely used in elliptic $R$-matrices:
 \beq\label{q901}
 \begin{array}{c}
  \displaystyle{
 T_a=T_{a_1 a_2}=\exp\left(\frac{\pi\imath}{{ N}}\,a_1
 a_2\right)Q^{a_1}\Lambda^{a_2}\,,\quad
 a=(a_1,a_2)\in\mZ_{ N}\times\mZ_{ N}\,,
 }
 \end{array}
 \eq
 where
 \beq\label{q902}
 \begin{array}{c}
  \displaystyle{
Q_{kl}=\delta_{kl}\exp(\frac{2\pi
 \imath}{{ N}}k)\,,\ \ \
 \Lambda_{kl}=\delta_{k-l+1=0\,{\hbox{\tiny{mod}}}\,
 { N}}\,,\quad Q^{ N}=\Lambda^{ N}=1_{{ N}\times { N}}\,.
 }
 \end{array}
 \eq
These are the generators of the finite dimensional representation of
Heisenberg group
 \beq\label{q903}
 \begin{array}{c}
  \displaystyle{
 \Lambda^{a_2} Q^{a_1}=\exp\left(\frac{2\pi\imath}{{ N}}\,a_1
 a_2\right)Q^{a_1}\Lambda^{a_2}\,.
 }
 \end{array}
 \eq
 Then for the product of basis matrices we have
  \beq\label{q904}
 \begin{array}{c}
  \displaystyle{
T_\al T_\be=\kappa_{\al,\be} T_{\al+\be}\,,\ \ \
\kappa_{\al,\be}=\exp\left(\frac{\pi \imath}{{ N}}(\be_1
\al_2-\be_2\al_1)\right)\,,
 }
 \end{array}
 \eq
 where $\al+\be=(\al_1+\be_1,\al_2+\be_2)$.
Therefore
  \beq\label{q905}
 \begin{array}{c}
  \displaystyle{
\tr(T_\al T_\be)={ N}\delta_{\al,-\be}\,,
 }
 \end{array}
 \eq
 The permutation operator
takes the following form in this basis:
  \beq\label{q906}
 \begin{array}{c}
  \displaystyle{
P_{12}=\sum\limits_{i,j=1}^{ N} { E}_{ij}\otimes {
E}_{ji}=\frac{1}{{ N}}\sum\limits_{\al\in\,\mZ_{ N} \times \mZ_{ N}}
T_\al \otimes T_{-\al}\,,
 }
 \end{array}
 \eq
where ${ E}_{ij}$ is the standard basis in ${\rm Mat}_{ N}$.

\paragraph{Theta functions.}
 The Riemann theta-functions with characteristics
 \beq\label{q907}
 \begin{array}{c}
  \displaystyle{
\theta{\left[\begin{array}{c}
a\\
b
\end{array}
\right]}(z|\, \tau ) =\sum_{j\in \mathbb\, Z}
\exp\left(2\pi\imath(j+a)^2\frac\tau2+2\pi\imath(j+a)(z+b)\right)\,,\quad
a\,,b\in\frac{1}{N}\,\mZ\,.
 }
 \end{array}
 \eq
are defined on elliptic curve
$\Sigma_\tau={\mC^2}/(\mZ\oplus\tau\mZ)$ with moduli $\tau$
(Im$\tau>0$). They behave on the lattice $\mZ\oplus\tau\mZ$) as
follows:
 \beq\label{q908}
 \begin{array}{c}
  \displaystyle{
\theta{\left[\begin{array}{c}
a\\
b
\end{array}
\right]}(z+1|\,\tau )=\exp(2\pi\imath a)\,
\theta{\left[\begin{array}{c}
a\\
b
\end{array}
\right]}(z |\,\tau )\,,
 }
 \end{array}
 \eq
 \beq\label{q909}
 \begin{array}{c}
  \displaystyle{
\theta{\left[\begin{array}{c}
a\\
b
\end{array}
\right]}(z+a'\tau|\,\tau ) =\exp\left(-2\pi\imath {a'}^2\frac\tau2
-2\pi\imath a'(z+b)\right) \theta{\left[\begin{array}{c}
a+a'\\
b
\end{array}
\right]}(z|\,\tau )\,.
 }
 \end{array}
 \eq
 We also use a shorthand notation for the odd theta function
 \beq\label{q910}
 \begin{array}{c}
  \displaystyle{
\vth(z|\,\tau)\equiv\vth(z)\equiv\theta{\left[\begin{array}{c}
1/2\\
1/2
\end{array}
\right]}(z|\, \tau )\,.
 }
 \end{array}
 \eq

\paragraph{Kronecker and Eisenstein functions.}
 The Kronecker function is defined in terms
of (\ref{q910}):
  \beq\label{q911}
  \begin{array}{l}
  \displaystyle{
 \phi(\eta,z)=\frac{\vth'(0)\vth(z+q)}{\vth(z)\vth(q)}
 }
 \end{array}
 \eq
The first Eisenstein and the second Eisenstein functions
  \beq\label{q912}
  \begin{array}{c}
  \displaystyle{
E_1(z)=\frac{\vth'(z)}{\vth(z)}\,,\quad\quad
E_2(z)=-\p_zE_1(z)=\wp(z)-\frac{1}{3}\frac{\vth'''(0)}{\vth'(0)}\,,
 }
 \end{array}
 \eq
where $\wp(z)$ is the Weierstrass  $\wp$-function. The function
$E_2(z)$ is double-periodic on the lattice $\mC/\mZ+\tau\mZ$, while
for the first Eisenstein and the Kronecker functions we have:
  \beq\label{q9121}
  \begin{array}{c}
  \displaystyle{
 E_1(z+1)=E_1(z)\,,\quad\quad E_1(z+\tau)=E_1(z)-2\pi\imath\,,
 }
 \end{array}
 \eq
  \beq\label{q9122}
  \begin{array}{c}
  \displaystyle{
 \phi(z+1,w)=\phi(z,w)\,,\quad\quad \phi(z+\tau,w)=e^{-2\pi\imath
 w}\phi(z,w)\,.
 }
 \end{array}
 \eq
The following set of functions numerated by
$a=(a_1,a_2)\in\mZ_N\times\mZ_N$ (as in (\ref{q901})) is also used
 \beq\label{q913}
 \begin{array}{c}
  \displaystyle{
 \vf_a(z,\om_a)=\exp(2\pi\imath\frac{a_2}{N}\,z)\,\phi(z,\om_a+\hbar)\,,\quad
 \om_a=\frac{a_1+a_2\tau}{N}\,.
 }
 \end{array}
 \eq
%

\paragraph{Genus one Fay trisecant identity} is as follows:
  \beq\label{q914}
  \begin{array}{c}
  \displaystyle{
\phi(\hbar,z)\phi(\eta,w)=\phi(\hbar-\eta,z)\phi(\eta,z+w)+\phi(\eta-\hbar,w)\phi(\hbar,z+w)
 }
 \end{array}
 \eq
 Its degenerations:
  \beq\label{q915}
  \begin{array}{c}
  \displaystyle{
 \phi(\eta,z)\phi(\eta,w)=\phi(\eta,z+w)(E_1(\eta)+E_1(z)+E_1(w)-E_1(z+w+\eta))\,,
 }
 \end{array}
 \eq
  \beq\label{q916}
  \begin{array}{c}
  \displaystyle{
 \phi(\hbar,z)\phi(\hbar,-z)=\wp(\hbar)-\wp(z)=E_2(\hbar)-E_2(z)\,.
 }
 \end{array}
 \eq

For the derivative of the Kronecker function with respect to the
second argument we keep notation
 \beq\label{q917}
 \begin{array}{c}
  \displaystyle{
f(z,q)=\p_q\phi(z,q)=\phi(z,q)(E_1(z+q)-E_1(q))\,.
 }
 \end{array}
 \eq
It satisfies identities:
 \beq\label{q918}
 \begin{array}{c}
  \displaystyle{
\phi(z,q)f(z,-q)-f(z,q)\phi(z,-q)=\wp'(q)\,,
 }
 \end{array}
 \eq
  \beq\label{q919}
  \begin{array}{c}
  \displaystyle{
\phi(z,q_{ab})f(z,q_{bc})-f(z,q_{ab})\phi(z,q_{bc})=
\phi(z,q_{ac})(\wp(q_{ab})-\wp(q_{bc}))\,.
 }
 \end{array}
 \eq
 Due to the local expansion near $z=0$
  \beq\label{q920}
  \begin{array}{c}
  \displaystyle{
\phi(z,q)=\frac{1}{z}+E_1(q)+\frac{1}{2}\left(E_1^2(q)-\wp(q)\right)+O(z^2)
 }
 \end{array}
 \eq
 we also have
  \beq\label{q921}
  \begin{array}{c}
  \displaystyle{
f(0,q)=-E_2(q)\,.
 }
 \end{array}
 \eq
 %
%

\paragraph{Heat equation.}
For the theta functions (\ref{q907}) the following relation holds
  \beq\label{q9291}
  \begin{array}{c}
  \displaystyle{
4\pi\imath\p_\tau \theta{\left[\begin{array}{c}
a\\
b
\end{array}
\right]}(z|\, \tau )=\p^2_z\theta{\left[\begin{array}{c}
a\\
b
\end{array}
\right]}(z|\, \tau )\,.
 }
 \end{array}
 \eq
In particular, it is true for $\vth(z)$ (\ref{q910}). Then using the
definitions (\ref{q911})-(\ref{q912}) we can get
  \beq\label{q9292}
  \begin{array}{c}
  \displaystyle{
 2\pi\imath\p_\tau\phi(z,q)=\p_z\p_q\phi(z,q)
 }
 \end{array}
 \eq
and
  \beq\label{q9293}
  \begin{array}{c}
  \displaystyle{
 2\pi\imath\p_\tau\log\vth(z)=\frac12(E_1^2(z)-E_2(z))\,.
 }
 \end{array}
 \eq
 %

\paragraph{Identities.}

 \beq\label{q9541}
 \begin{array}{c}
  \displaystyle{
 (E_1(x)+E_1(y)+E_1(-x-y))^2=\wp(x)+\wp(y)+\wp(x+y)=
 }
 \\  \ \\
  \displaystyle{
 \stackrel{(\ref{q912})}{=}E_2(x)+E_2(y)+E_2(x+y)+\frac{\vth'''(0)}{\vth'(0)}
 }
 \end{array}
 \eq

  \beq\label{q950}
  \begin{array}{l}
  \displaystyle{
\frac{\vth(\hbar)}{\vth'(0)}\,\sum\limits_k g_{ik}(z,q)\,
\phi(z,q_{kj}+\hbar)=g_{ij}(z+N\hbar,q) \prod\limits_{m\neq
j}\frac{\vth(q_{mj})}{\vth(q_{mj}+\hbar)}\,.
 }
 \end{array}
 \eq

For
 \beq\label{q951}
 \begin{array}{c}
  \displaystyle{
X_{ij}(x_j)=
 \vth\left[  \begin{array}{c}
 \frac12-\frac{i}{N} \\ \frac N2
 \end{array} \right] \left(Nx_j\left.\right|N\tau\right)
 }
 \end{array}
 \eq
 we have
 \beq\label{q952}
 \begin{array}{c}
  \displaystyle{
 \det X=C_N(\tau)\,\vth(\sum\limits_{k=1}^N
 x_k)\prod\limits_{i<j}\vth(x_j-x_i)\,,\quad\quad
 C_N(\tau)=\frac{(-1)^{N-1}}{(\imath\eta(\tau))^{\frac{(N-1)(N-2)}{2}}}\,,
 }
 \end{array}
 \eq
where $\eta(\tau)$ is the Dedekind eta-function:
 \beq\label{q953}
 \begin{array}{c}
  \displaystyle{
 \eta(\tau)=e^{\frac{\pi\imath\tau}{12}}\prod\limits_{k=1}^\infty (1-e^{2\pi\imath\tau
 k})\,.
 }
 \end{array}
 \eq
Then for the matrix (\ref{q53})
 \beq\label{q954}
 \begin{array}{c}
  \displaystyle{
 \det\Xi(z,q)=C_N(\tau)\,\vth(z)\prod\limits_{i<j}\vth(q_i-q_j)\,.
 }
 \end{array}
 \eq

%

 \section{Appendix B}\label{AppB}
\def\theequation{B.\arabic{equation}}
\setcounter{equation}{0}

\paragraph{Proof of formula (\ref{q509}).}
To prove  (\ref{q509}) we need to show that
 \beq\label{q955}
  J = D^{0}V^{-1}\tilde{C}V(D^{0})^{-1} = \left(
\begin{array}{cc}
    \frac{\delta_{ij}}{2q_{i}} & -\frac{\delta_{ij}}{2q_{i}}\\
    \frac{\delta_{ij}}{2q_{i}} & -\frac{\delta_{ij}}{2q_{i}}
 \end{array}\right).
 \eq
  The proof of (\ref{q955}) is a direct evaluation, which uses  explicit
form of the inverse Vandermonde matrix:
 \beq\label{q956}
 \begin{array}{c}
    \displaystyle{
        J_{ij}=(D^{0}V^{-1}\tilde{C}V(D^{0})^{-1})_{ij}=D^{0}_{i\alpha}V^{-1}_{\alpha  \beta}\tilde{C}_{ \beta \gamma}
        V_{\gamma \nu}(D^{0})^{-1}_{\nu j}=
    }
    \\ \ \\
    \displaystyle{
        =\sum\limits_{\gamma - {odd}}\frac{D^{0}_{i}}{D^{0}_{j}}V^{-1}_{i,\gamma+1}V_{\gamma,j}=
        \frac{D^{0}_{i}}{q_{j}D^{0}_{j}}\sum\limits_{\gamma - {even} }V^{-1}_{i\gamma}V_{\gamma
        j}\,.
    }
 \end{array}
  \eq
To see how the matrix $J_{ij}$ changes under substitutions $i
\rightarrow i+N$ and $j \rightarrow j+N$ consider the changes of its
factors: $D^{0}_{j} \rightarrow -D^{0}_{j}$, $(D^{0})^{-1}_{i}
\rightarrow -(D^{0})^{-1}_{i}$ $q_{i} \rightarrow -q_{i}$,
$V_{\gamma j} \rightarrow -V_{\gamma j}$, $V^{-1}_{i \gamma}
\rightarrow -V^{-1}_{ i \gamma}$. The penultimate relation holds
true because since the summation goes over odd $\gamma$. Therefore,
$J_{ij}$ does not change the sign under the substitution $i
\rightarrow i+N$, and  $J_{ij}$ changes the sign under $j
\rightarrow j+N$. Thus the matrix $J$ has the form:
 \begin{equation}
\label{q957} J=\left( \begin{array}{cc}
\tilde{J}&-\tilde{J}\\
\tilde{J}&-\tilde{J}
 \end{array}\right).
 \end{equation}
Further, we will consider $1 \leq i,j \leq N$, since this is
sufficient to determine matrix $J$.
 \beq\label{q958}
 \begin{array}{c}
    \displaystyle{
        \sum\limits_{\gamma-even}V^{-1}_{i\ga}V_{\ga j} = \sum\limits_{\ga - odd}V^{-1}_{i,\ga+1}V_{\ga+1,j} =
    }
    \\ \ \\
    \displaystyle{
        =\sum\limits_{\ga-odd}q_{j}^{\ga}\frac{1}{\ga!}\partial_{\rho}^{(\ga)}\left.\left[\frac{(\rho+q_{i})
 \prod\limits_{s\neq i}(\rho-q_{s})(\rho+q_{s})}{2q_{i}
 \prod\limits_{s \neq i}(q_{i}-q_{s})(q_{i}+q_{s})}\right]\right|_{\rho=0} =
    }
    \\ \ \\
    \displaystyle{
        =\sum\limits_{\ga-odd}q_{j}^{\ga}\frac{1}{(\ga-1)!}\partial_{\rho}^{(\ga-1)}
        \left.\left[\frac{\prod\limits_{s\neq i}(\rho-q_{s})(\rho+q_{s})}{2q_{i}
        \prod\limits_{s \neq i}(q_{i}-q_{s})(q_{i}+q_{s})}\right]\right|_{\rho=0} =
    }
    \\ \ \\
    \displaystyle{
        =\frac{q_{j}}{q_{i}}\sum\limits_{\ga-odd}q_{j}^{\ga-1}\frac{1}{(\ga-1)!}\partial_{\rho}^{(\ga-1)}
        \left.\left[\frac{\prod\limits_{s\neq i}(\rho-q_{s})(\rho+q_{s})}{2
        \prod\limits_{s \neq i}(q_{i}-q_{s})(q_{i}+q_{s})}\right]\right|_{\rho=0} =
    }
    \\ \ \\
    \displaystyle{
        =\frac{q_{j}}{q_{i}}\sum\limits_{\ga-odd}q_{j}^{\ga-1}\frac{1}{(\ga-1)!}\partial_{\rho}^{(\ga-1)}
        \left.\left[\frac{(\rho+q_{i})\prod\limits_{s\neq i}(\rho-q_{s})(\rho+q_{s})}{2q_{i}
        \prod\limits_{s \neq i}(q_{i}-q_{s})(q_{i}+q_{s})}\right]\right|_{\rho=0} =
        \frac{q_{j}}{q_{i}}\sum\limits_{\ga -odd}V^{-1}_{i \ga}V_{\ga
        j}\,.
    }
 \end{array} \eq
Using this relation we obtain:
  \beq\label{q959} \sum\limits_{\ga
-even}V^{-1}_{i \ga}V_{\ga j} = \frac{\delta_{ij}}{2}
 \eq
  and, therefore
 \beq\label{q960}
 \begin{array}{c}
      \displaystyle{
    \tilde{J}_{ij}=\frac{D^{0}_{i}}{q_{j}D^{0}_{j}}\frac{\delta_{ij}}{2} =
    \frac{\delta_{ij}}{2q_{j}}\,.
    }
 \end{array}
  \eq
In this way the formula (\ref{q509}) is proved.


\vskip4mm\noindent \textbf{Proof of formula (\ref{q515})}: First,
determine the structure of matrix:
 \beq\label{q961}
 \begin{array}{c}
     \displaystyle{
    G_{ij}= (D^{0}V^{-1}(C+\tilde{C})V(D^{0})^{-1})_{ij}=\frac{D^{0}_{i}}{D^{0}_{j}}\sum\limits_{\gamma =1}^{2N}
    V^{-1}_{i,\ga+1}(C+\tilde{C})_{\ga+1,\ga}V_{\ga j}\,.
    }
 \end{array}
  \eq
Consider $1 \leq i,j \leq N$. Let us find out how its matrix
elements change under the substitutions $i \rightarrow i+N$ and $j
\rightarrow j+N$:
 \beq\label{q962}
 \begin{array}{c}
    D^{0}_{i} \rightarrow D^{0}_{i},\:\:
    D^{0}_{j} \rightarrow D^{0}_{j},\:\:
    V_{\ga j} \rightarrow (-1)^{\ga-1}V_{\ga j},\:\:
    V^{-1}_{i,\ga+1} \rightarrow (-1)^{\ga}V^{-1}_{i,\ga+1}\,.
 \end{array}
  \eq
 Therefore, $G_{ij} \rightarrow -G_{ij}$. Similar properties are valid  for $1 \leq i \leq N$,
$N+1 \leq j \leq 2N$ and the substitutions $i \rightarrow i+N$, $j
\rightarrow j-N$. Thus, we get $G_{ij} \rightarrow -G_{ij}$. Then
consider the case $i=j=2N+1$:
 \beq\label{q963} G_{2N+1,2N+1} =
\sum\limits_{\ga=1}^{2N}V^{-1}_{2N+1,\ga+1}(C+\tilde{C})_{\ga+1,
\ga}V_{\ga, 2N+1} = V^{-1}_{2N+1,2}(C+\tilde{C})_{2,1} = 0
  \eq
 and the case $j=2N+1$, $1\leq i \leq N$:
 \beq\label{q964}
 \begin{array}{c}
    \displaystyle{
        G_{i,2N+1}=\frac{D^{0}{i}}{D^{0}_{2N+1}}\sum\limits_{\ga=1}^{2N}V^{-1}_{i,\ga+1}(C+\tilde{C})_{\ga+1,\ga}V_{\ga,2N+1} =
        2\frac{D^{0}_{i}}{D^{0}_{2N+1}}V^{-1}_{i2}=
    }
    \\ \ \\
    \displaystyle{
        =2\frac{D^{0}_{i}}{D^{0}_{2N+1}} \partial_{\rho}\Big[\left.\frac{\rho(\rho+q_{i})
        \prod\limits_{s\neq i}(\rho^{2}-q_{s}^{2})}{2q_{i}^{2}\prod\limits_{s \neq i}(q^{2}_{i}-q_{s}^{2})}\Big]\right|_{\rho=0}
        =-\frac{D^{0}_{i}}{D^{0}_{2N+1}}\frac{1}{q_{i}^{3}}\frac{\prod\limits_{s}(-q_{s})^{2}}{\prod\limits_{s \neq i}(q_{i}^{2}-q_{s}^{2})}=
-\frac{\sqrt{2}}{q_{i}}\,.
    }
 \end{array}
  \eq
 Similarly, for $N+1 \leq i \leq 2N$ we get:
  \beq\label{q965}
 \begin{array}{c}
     \displaystyle{
    G_{i,2N+1}=\frac{\sqrt{2}}{q_{i-N}}\,.
    }
 \end{array}
  \eq
 Calculate $G_{2N+1,j}$ for $1 \leq j \leq N$:
  \beq\label{q966}
 \begin{array}{c}
    \displaystyle{
        G_{2N+1,j} = \frac{D^{0}_{2N+1}}{D^{0}_{j}}\sum\limits_{\ga=1}^{2N}V^{-1}_{2N+1,\ga+1}(C+\tilde{C})_{\ga+1,\ga}V_{\ga,j} =
    }
    \\ \ \\
    \displaystyle{
        = \frac{D^{0}_{2N+1}}{D^{0}_{j}}\sum\limits_{\ga - even}^{2N}\left.
        \frac{q_{j}^{\ga-1}}{(\ga-1)!}\partial_{\rho}^{(\ga)}\Big[
        \frac{\prod\limits_{s=1}^{N}(\rho^{2}-q_{s}^{2})}{D^{0}_{2N+1}}\Big]\right|_{\rho=0}=
    }
    \\ \ \\
    \displaystyle{
        =\frac{1}{D^{0}_{j}}\sum\limits_{\ga-even}q_{j}^{\ga-1}\frac{1}{(\ga-1)!}\partial_{\rho}^{(\ga)}\Big[(\rho^{2}-q_{j}^{2})
        \left.\prod\limits_{s\neq j}^{N}(\rho^{2}-q_{s}^{2})\Big]\right|_{\rho=0}=
    }
 \end{array}
  \eq
 $$
 \begin{array}{c}
    \displaystyle{
        =\frac{1}{D^{0}_{j}}\sum\limits_{\ga -even}q_{j}^{\ga-1}\left.\Big[\frac{\ga}{(\ga-2)!}\partial_{\rho}^{(\ga-2)}
        \big[\prod\limits_{s \neq j}(\rho^{2}-q_{s}^{2})\big]\right|_{\rho=0} -
    }
        \\ \ \\
    \displaystyle{
        -\frac{q_{j}^{2}}{(\ga-1)!}\partial_{\rho}^{(\ga)}\big[\left.\prod\limits_{s \neq j}(\rho^{2}-q_{s}^{2})\big]\right|_{\rho=0} \Big]=
  }
    \\ \ \\
    \displaystyle{
        =\frac{1}{D^{0}_{j}}\sum\limits_{\ga -even}q_{j}^{\ga-1}\Big[\frac{\ga}{q_{j}(\ga-2)!}\partial_{\rho}^{(\ga-2)}\big[(\rho+q_{j})
        \left.\prod\limits_{s \neq j}(\rho^{2}-q_{s}^{2})\big]\right|_{\rho=0} -
    }
        \\ \ \\
    \displaystyle{
        -\frac{q_{j}}{(\ga-1)!}\partial_{\rho}^{(\ga)}\big[(\rho+q_{j})\left.\prod\limits_{s \neq j}(\rho^{2}-q_{s}^{2})\big]\right|_{\rho=0}
        \Big]\,.
    }
 \end{array}
 $$
  Therefore,
 \beq\label{q967}
 \begin{array}{c}
    \displaystyle{
       G_{2N+1,j} =\frac{\sqrt{2}}{q_{j}}\left(\sum\limits_{\ga-even}q_{j}^{\ga-2}\frac{\ga}{(\ga-2)!}
        \partial_{\rho}^{(\ga-2)}\Big[ \left.\frac{(\rho+q_{j})\prod\limits_{s \neq j}(\rho^{2}-q_{j}^{2})}{2q_{j}
        \prod\limits_{s \neq
        j}(q_{j}^{2}-q_{s}^{2})}\Big] \right|_{\rho=0} \right)-
    }
    \\ \ \\
    \displaystyle{
        -\frac{\sqrt{2}}{q_{j}}\left(q_{j}^{\ga}\frac{\ga}{\ga!}\partial_{\rho}^{(\ga)}\Bigl[\left.\frac{(\rho+q_{j})
        \prod\limits_{s \neq j}(\rho^{2}-q_{s}^{2})}{2q_{j}\prod\limits_{s \neq j}(q_{j}^{2}-q_{s}^{2})} \Bigr]\right|_{\rho=0}\right) =
    }

 \end{array}
  \eq
$$
 \begin{array}{c}
    \displaystyle{
        =\frac{\sqrt{2}}{q_{j}}\Big(\sum\limits_{\ga -even}^{2N}\ga (V^{D})^{-1}_{j,\ga-1}V^{D}_{\ga-1,j} -
        \sum\limits_{\ga-even}^{2N-2}\ga (V^{D})^{-1}_{j,\ga+1}V^{D}_{\ga+1,j}  \Big)=
    }
    \\ \ \\
    \displaystyle{
        =\frac{2\sqrt{2}}{q_{j}}\sum\limits_{\ga-even}^{2N}(V^{D})^{-1}_{j,\ga-1}V^{D}_{\ga-1,j}=
        \frac{2\sqrt{2}}{q_{j}}\frac{\delta_{jj}}{2} =
        \frac{\sqrt{2}}{q_{j}}\,.
    }
 \end{array}
 $$
In the last equalities the result from the proof of formula
(\ref{q509}) was used.
 Under the substitution $j \rightarrow j+N$: $D^{0}_{j}\rightarrow
D^{0}_{j}$, $V_{\ga j} \rightarrow (-1)^{\gamma-1}V_{\ga j}$. Taking
into account that the sum goes over even $\gamma$ we obtain
$G_{2N+1,j} \rightarrow -G_{2N+1,j}$. Thus, we proved that the Lax
matrix is of the form:
 \begin{equation}
\label{q968} L(m_{2},0,\sqrt{2}m_{2})=\left( \begin{array}{ccc}
A&B&C\\
-B&-A&-C\\
-C^{T}&C^{T}&0\\
 \end{array}\right)\\,\:\:\:
C_{i}=\frac{\sqrt{2} m_{2}}{q_{i}}
 \end{equation}
Then, for the last two blocks  ($1 \leq i \leq N$, $N+1 \leq j \leq
2N$) we have:
  \beq\label{q969}
 \begin{array}{c}
    \displaystyle{
        B_{i,j-N}=-m_{2}\frac{D^{0}_{i}}{D^{0}_{j}}\sum\limits_{\ga=1}^{2N}V^{-1}_{i,\ga+1}(C+\tilde{C})_{\ga+1,\ga}V_{\ga,j} =
    }
    \\ \ \\
    \displaystyle{
        =-m_{2}\frac{D^{0}_{i}}{D^{0}_{j}}
        \sum\limits_{\ga=1}^{2N}(-q_{j-N})^{\ga-1}(C+\tilde{C})_{\ga+1,\ga}\frac{1}{\ga!}
        \left.\partial_{\rho}^{(\ga)}\Bigl[\frac{\rho(\rho+q_{i})\prod\limits_{s \neq i}(\rho^{2}-q_{s}^{2})}{2q_{i}^{2}
        \prod\limits_{s \neq i}(q_{i}^{2}-q_{s}^{2})}\Bigr]\right|_{\rho=0}=
    }
    \\ \ \\
    \displaystyle{
        =-m_{2}\frac{D^{0}_{i}}{D^{0}_{j}}\sum\limits_{\ga=1}^{2N}(-q_{j-N})^{\ga-1}(C+\tilde{C})_{\ga+1,\ga}\frac{1}{(\ga-1)!}
        \left.\partial_{\rho}^{(\ga-1)}\Bigl[\frac{(\rho+q_{i})
        \prod\limits_{s \neq i}(\rho^{2}-q_{s}^{2})}{2q_{i}^{2}\prod\limits_{s \neq i}(q_{i}^{2}-q_{s}^{2})}\Bigr]\right|_{\rho=0}=
    }
 \end{array}
  \eq
 $$
 \begin{array}{c}
    \displaystyle{
       = m_{2}\frac{q_{i}}{q_{j-N}}\frac{D^{0D}_{i}}{D^{0D}_{j}}\frac{1}{q_{i}}
        \left[
        \sum\limits_{\ga=1}^{2N}(-q_{j-N})^{\ga-1}((C+\tilde{C})_{\ga+1,\ga}-2)\frac{1}{(\ga-1)!}\times\right.
    }
    \\ \ \\
    \displaystyle{
    \left.\times\partial_{\rho}^{(\ga-1)}\Big[\left.\frac{(\rho+q_{i})
        \prod\limits_{s \neq i}(\rho^{2}-q_{s}^{2})}{2q_{i}\prod\limits_{s \neq i}(q_{i}^{2}-q_{s}^{2})}\Bigr]\right|_{\rho=0}
        +2\sum\limits_{\ga=1}^{2N}(V^{D})^{-1}_{i\ga}(V^{D}_{\ga j})^{-1}  \right]=
    }
     \end{array}
  $$
 $$
 \begin{array}{c}
    \displaystyle{
        =-m_{2}\frac{D^{0D}_{i}}{D^{0D}_{j}}
        \sum\limits_{\ga=2}^{2N}(-q_{j-N})^{\ga-2}(C_{\ga+1,\ga}+\tilde{C}_{\ga+1,\ga}-2)(V^{D})^{-1}_{i\ga} =
    }
    \\ \ \\
    \displaystyle{
        =-m_{2}\frac{D^{0D}_{i}}{D^{0D}_{j}}
        \sum\limits_{\ga=1}^{2N-1}(-q_{j-N})^{\ga-1}(C_{\ga+1,\ga}-\tilde{C}_{\ga+1,\ga})(V^{D})^{-1}_{i,\ga+1}=B^{D}_{ij}\,.
    }
 \end{array}
  $$
  Notice that the term $2 \cdot \sum\limits_{1\leq \gamma \leq
2N}(V^{D})^{-1}_{i\ga}V^{D}_{\ga j}= 2\delta_{ij}$ vanishes since $i
\neq j$ in our case. This gives the block $B$ from the Lax matrix in
$D_{N}$ case. The last one block corresponding to $1 \leq i \leq N$,
$1 \leq j \leq N$ is evaluated in a similar way,  apart from  the
term
 \beq\label{q970} 2\sum\limits_{1\leq \gamma \leq
2N}V^{D}_{i\gamma}(V^{D})^{-1}_{\gamma j}=2\delta_{ij}
  \eq
which does not vanish. Therefore, we get
 \beq\label{q971}
 \begin{array}{c}
     \displaystyle{
    A_{ij}=A^{D}_{ij}-m_{2}\frac{2\delta_{ij}}{q_{i}}\,.
    }
 \end{array}
  \eq
This finishes the proof of (\ref{q515}).

\vskip2mm

\noindent {\bf Acknowledgments.} The work was performed at the
Steklov Mathematical Institute of Russian Academy of Sciences,
Moscow. This work is supported by the Russian Science Foundation
under grant 14-50-00005.A. Zotov is a Young Russian Mathematics
award winner and would like to thank its sponsors and jury.

\begin{small}

\end{small}

\end{document}